\documentstyle[art10]{article}
\input{psfig}
\newcommand{\dir}{.}
\newcommand{\fig}[3]
{
\begin{center}
     \noindent
     \unitlength=1mm
     \begin{picture}(#2,#3)
     \put(0,0){
       \psfig{figure=\dir/#1,width=#2mm,height=#3mm}
     }
     \end{picture}
   \noindent
\end{center}
}

\begin{document}
\baselineskip=12pt
\setcounter{page}{1}

\noindent
{\Large\bf
Self-consistent field theories for complex fluids
}

\vspace{0.5cm}

\begin{center}
F. Schmid \\
{\em Institut f\"ur Physik, Universit\"at Mainz, D-55099 Mainz, FRG}
\end{center}

\begin{quote}
{\bf Abstract.}
Recent developments in off-lattice self-consistent field theories for 
inhomogeneous complex fluids are reviewed. Particular emphasis is given
to the treatment of intermolecular interactions and compressibility, to
the role of fluctuations, and to the discussion of the coarse-graining 
length which is inherent to the theory. Valuable insight can be gained from
the comparison of self-consistent field calculations with 
Monte Carlo simulations. Finally, some applications of the theory to
orientational properties of polymers and copolymers at interfaces, and to 
the phase behavior of amphiphiles at surfaces, are presented.
\end{quote}

\section{Introduction}

Surfaces and interfaces in supramolecular fluids are a topic of interest
in various contexts of materials science and soft condensed matter physics.

On the one hand, the surface structure of materials determines many surface 
properties (adhesive and wetting properties, optical properties etc.) and 
thus plays an important role for many applications, such as lubricants, 
coatings, thin films and membranes\cite{WU1,GARBASSI}. 
If the topography and chemistry of surfaces can be controlled, this can
be exploited to taylor surfaces for specific purposes\cite{MAYES},
{\em e.g}, surfaces which anchor liquid crystals in a well-defined way for
liquid crystal cells\cite{GEARY}, intelligent surfaces\cite{WILLSON} etc.

On the other hand, interfacial properties often influence the bulk 
behavior of materials in a substantial way, since many 
commonly used materials are inhomogeneous on a microscopic scale.
An important class of substances, where that is the case,
are melts of polymers which contain 
monomers of different type, {\em e.g.}, polymer blends or copolymer melts.
This is because different organic molecules are often slightly 
incompatible\cite{HILDEBRAND}. In homopolymer alloys ({\em e.g.}, composite 
matrix materials such as rubber toughened plastics), the small relative 
repulsion of the monomers is amplified by the large number of monomers 
in the macromolecules, and completely dominates the entropy of mixing, 
which is proportional to the number of molecules.
As a result, polymers of different type are usually immiscible at temperatures 
of practical interest. They consist microscopically of numerous 
finely dispersed droplets of one component in the other, and the interfaces 
between these essentially unmixed phases largely govern the material 
properties\cite{WU2,WOOL,RALF}.

One way to prevent the components from demixing is to chemically bind them
to each other, thus forming block copolymers. Even though macroscopic phase
separation is then inhibited, the chains still have a tendency to rearrange 
themselves so as to allow like monomers to pack next to each other. Depending
on the temperature, block size etc., this may lead to ``microphase
separation'', {\em i.e}, the formation of ordered mesoscopic structures -- 
of lamellae, ordered micelle arrays, bicontinuous structures\cite{BATES1} --
which can be viewed as ordered arrays of interfaces in a sense.
The properties of those materials strongly depend on the topology and can be
quite unusual\cite{BATES3,TEMPLIN}.
Copolymers are also used as effective compatibilizers in homopolymer blends. 
Added in small amounts, they reduce the interfacial tension between the
homopolymer phases\cite{ANASTASIADIS}, and dynamically prevent the coalescence 
of droplets\cite{BECK,MILNER}. Increasing the copolymer concentration results
again in the formation of mesoscopically structured microphase-separated
phases\cite{phdexp,phdth}.

Other prominent examples for inhomogeneous complex fluids are
self aggregating amphiphilic systems, {\em e.g.}, mixtures of oil,
water and soap, or lipid-water systems. At high enough amphiphile
concentration, they build ordered structures which are very similar
to the above mentioned copolymer mesophases\cite{LUZZATI,GOMPPERB}. The study
of lipids in water is particularly interesting because of the
close connection to biological physics. Indeed, bilayers of lipid molecules
are a major ingredient of biological membranes and thus omnipresent
in any living organism.

Complex fluids are typically made of chemically complicated, large
molecules. They are usually characterized by a variety of competing length 
scales -- in the case of polymer melts the size of the monomers vs. the
extension of whole molecules, in the case of amphiphilic systems 
also the correlation within and between the self aggregated structures
-- and by the presence of additional, conformation and/or orientational, 
degrees of freedom. A full treatment in atomistic detail is in most cases 
out of reach of today's supercomputers. On the other hand, the apparent 
complexity of the systems actually contributes to a simplification of the 
physics on a coarse-grained level. Due to the large number of
possible interactions between molecules, microscopic details average out
to a large extent. A few characteristic attributes of the molecules are
often responsible for the main features of a substance. For
example, the properties of polymeric materials are determined mostly
by the chain character and the flexibility of the polymers; amphiphilic
systems are characterized by their affinity of the amphiphiles to both polar 
and nonpolar environments, and by their orientation at interfaces.
This motivates the study of idealized simplified models, which account only 
for the main properties of the molecules and absorb the microscopic details 
in a few, effective parameters\cite{FLORYB,DEGENNESB,FREEDB,BINDERB,BINDERR}.
A second important point is that dense macromolecular systems are
often unusually well described by mean field approximations.
Since large molecules interact with many others, the effective interaction 
range in the limit of high molecular weight is very large, and according
to a simple Ginzburg type argument, the critical region in which 
concentration fluctuations become important is very small as 
a result\cite{ginzburg}.

Mean field theories are thus widely used to describe inhomogeneous 
macromolecular systems. Many simple and largely successful theories are 
based on Ginzburg-Landau type functionals of the concentration or other
extensive quantities (fields), which characterize a system with respect to 
the problem of interest\cite{HOLYSTR}. The interactions between molecules
are effectively incorporated into square gradient terms and sometimes 
contributions of higher order derivatives of the fields\cite{GOMPPERB}
The parameters of the model are either derived from a microscopic theory 
(as in the de Gennes--Flory--Huggins theory of inhomogeneous polymer
mixtures\cite{DEGENNESB}), or from comparison with experiment. 
Such a treatment is adequate as long as the interactions are short ranged,
the inhomogeneities are only weak and characterized by length
scales of the order of the molecular size. It becomes questionable
in systems which are inhomogeneous on smaller length scales -- which
is true for most examples cited above. A substantial amount of effort has 
therefore been put into developing more refined mean field theories, 
which explicitly account for the chain conformational structure
(lattice theories such as the Flory-Huggins 
theory\cite{FLORY1,FLORYB,FREED1,KIKUCHI}, 
density functional theories\cite{FREED2,WOODWARD,MCCOY,KIERLIK},
self-consistent field 
theories\cite{HELFAND1,HELFAND2,NOOLANDI1,SHULL1,SCHEUTJENS}),
or even for the local structure in polymer melts
({\em e.g.}, the lattice cluster theory by Freed and 
coworkers\cite{FREED3},
 or the P-RISM theory by Schweizer and Curro\cite{SCHWEIZERR}).
The more microscopic factors a theory incorporates, the more detailed
is the information that it can provide on local structure properties.
Unfortunately, the treatment also gets more and more involved. 
Theories with a higher level of coarse-graining have the advantage
of being somewhat more transparent and easier to handle. 
Moreover, the origins of given physical phenomena can sometimes be
tracked down more easily, if they can be discussed in terms of a highly 
idealized model. The optimal choice of a theory of course depends 
on the specific questions one wishes to address.

In this paper, we shall discuss one of the most powerful mean field tools 
in the study of macromolecular systems, the self-consistent field approach. 
Even though, strictly speaking, every mean field approximation involves
self-consistent fields, the name of self-consistent field (SCF) theory
is in the polymer community by tradition reserved to a certain type of 
approach, which goes back some time ago to Edwards \cite{EDWARDS} and
Helfand and Tagami\cite{HELFAND1}. In a nutshell, it can be characterized as 
follows: Polymers are described as random walks in a positional dependent
chemical potential, which depends in turn in some self-consistent way 
on the distribution of monomers. The basic idea is thus extremely 
simple, which explains the enormous appeal that it has had for polymer
scientists ever since it was first formulated. A particularly popular
version of the theory has been developed by Scheutjens and Fleer
for lattice models\cite{SCHEUTJENS} and applied to various problems related
to polymers at interfaces or surfaces \cite{FLEERB}. Lattice models allow
for an efficient study of systems with a planar geometry,
as long as one is not crucially interested in chain stiffness and local
orientation effects. The treatment of situations which involve 
strongly curved interfaces and local or global orientational order is 
more difficult. One way out of this dilemma is to employ more sophisticated
lattice models. For example, the bond fluctuation model of Carmesin and 
Kremer\cite{CARMESIN}, a popular lattice model in Monte Carlo 
simulations\cite{BINDERB,MARCUSR}, has recently been used for self-consistent 
field studies of lipid bilayers and polymer alloy systems\cite{MARCUS1}. 
On the other hand, the above mentioned difficulties are automatically 
eliminated if one works in continuous space. We shall restrict our discussion 
mostly to such off-lattice models in the following. 
The basic concepts and recent methodical advances shall be reviewed in the 
next section. In section 3, some applications shall be presented which 
illustrate the use of the model for the study of polymer interfaces and 
amphiphilic systems. We summarize and conclude in section 4.

\section{Theory}

\subsection{Basic concepts}

We consider a mixture of $n_j$ molecules of type $j=a,b,c \cdots$ in a
volume $V$, which are built from $N_j$ monomers of species 
$\alpha = A,B,C \cdots$. Here roman indices distinguish between molecule 
types, and greek indices denote monomer species. For simplicity, we shall also
assume that the monomers in a molecule build a linear sequence -- the 
generalization to different architectures is straightforward\cite{MARK1}. 
The molecules can then be represented by continuous space curves 
$\vec{R}(s)$, with $s$ varying between 0 and $N_j$. Polymer types $j$ 
differ from each other by their chain length $N_j$, and by the distributions 
of monomers $\alpha$ on the chains. The latter are conveniently described by 
functions $\gamma_{\alpha,j}(s)$, which are 1 for portions of the polymer 
occupied by the $\alpha$-monomers, and 0 otherwise. Hence one has
$\sum_{\alpha} \gamma_{\alpha,j}(s) = 1$ for all $j$ and $s$. 
In $A$-homopolymers, for example, $\gamma_{\alpha}(s)=\delta_{A,\alpha}$ for 
all $s \in [0,N_j]$, and in the case of symmetric $A:B$ diblock 
copolymers, $\gamma_{\alpha}(s) = \delta_{A,\alpha}$ for $s \in [0,N_j/2]$ 
and $\gamma_{\alpha}(s) = \delta_{B,\alpha}$ for $s \in [N_j/2,1]$. 

For a given configuration $\{ \vec{R}_{i_j}(\cdot) \}$, one can then define
a local monomer density operator 
\begin{equation}
\label{rho}
\widehat{\rho}_{\alpha}(\vec{r}) = \sum_j n_j \;
\widehat{\rho}^j_{\alpha}(\vec{r}) 
\end{equation}
\begin{displaymath}
\mbox{with}\qquad
\widehat{\rho}^j_{\alpha}(\vec{r})  =
\frac{1}{n_j} \sum_{i_j=1}^{n_j} \int_0^{N_j} ds \: 
\delta(\vec{r} - \vec{R}_{i_j}(s) )
\: \gamma_{\alpha,j}(s) .
\end{displaymath}
The interaction between monomers can be described by a coarse-grained
monomer free energy functional ${\cal V} \{ \widehat{\rho}_{\alpha} \}$.
In addition to the Boltzmann factor associated with this interaction
energy, chain conformations are distributed according to an intrinsic
statistical weight ${\cal P}_j\{\vec{R}(\cdot)\}$, which accounts for
the internal energy of the chain, and for the part of the configurational
entropy which stems from length scales smaller than the coarse-graining length.
In practice, polymers are most commonly modelled as gaussian 
chains\cite{HELFAND3},
\begin{equation}
\label{pgauss}
{\cal P}_j\{\vec{R}(\cdot)\} =
{\cal N} \; \exp\Big[
- \sum_{\alpha} \frac{3}{2 b_{\alpha}{}^2} 
\int_0^{N_j} \! ds \: \Big| \frac{d \vec{R}(s)}{ds} \Big|^2
\gamma_{\alpha,j}(s) \Big],
\end{equation}
with the statistical segment length of $\alpha$-monomers $b_{\alpha}$, and
the normalization factor ${\cal N}$ chosen such that
\mbox{$\int{\cal D}\{\vec{R}(\cdot)\}\: {\cal P}_j\{\vec{R}(\cdot)\}=V$}.
Here and in the following, all lengths are given in units of some microscopic 
length $V_0^{1/3}$ and hence dimensionless. The conformational free energy 
of the chains is thus approximated by a gaussian stretching energy with 
spring constant $3/2 b_{\alpha}{}^2$. Shorter or stiffer polymers are 
sometimes represented as ``wormlike'' chains of fixed contour length. 
Chain portions made of $\alpha$ monomers are then characterized by 
the monomer length $a_\alpha$ and the dimensionless stiffness $\eta_{\alpha}$, 
and the probability distribution functionals ${\cal P}_j$ are given 
by\cite{KRATKI}
\begin{equation}
\label{pworm}
{\cal P}_j\{\vec{R}(\cdot)\} =
{\cal N} \; \prod_s \delta(\vec{U}^2-1) \;
\exp\Big[
- \sum_{\alpha} \frac{\eta_{\alpha}}{2} 
\int_0^{N_j} \! ds \: \Big| \frac{d \vec{U}(s)}{ds} \Big|^2
\gamma_{\alpha,j}(s) \Big],
\end{equation}
where $\vec{U} = (d \vec{R}/ds)/a_{\alpha}$ is a dimensional tangent vector
constrained to unity by the delta function. This choice of a statistical
weight has the advantage that monomer orientations are well-defined and 
orientation dependent interactions between monomers can be introduced in a 
straightforward way. In the limit of small $\eta$, 
{\em i.e.}, very flexible chains, the wormlike or semiflexible chain 
statistics reduces to gaussian chain statistics\cite{MORSE}.
Of course, it is also possible to consider chains with mixed gaussian
and wormlike parts.

Alternatively, many polymer models describe molecules by discrete walks 
$\{ \vec{R}_s \}$ instead of continuous curves $\{ \vec{R}(s)\}$. 
The treatment is essentially the same, save that integrals $\int_0^{N_j}\!ds$ 
are replaced by sums $\sum_s$. In the case of freely jointed 
chains with fixed bond length $ l$,
for example, the statistical weight is 
\begin{equation}
\label{pjoint}
{\cal P}_j\{\vec{R}(\cdot)\} =
{\cal N} \; \prod_{s=1}^{N_j-1} 
\delta \Big( |\vec{R}_{s+1}-\vec{R}_s|-
l \;
\Big).
\end{equation}
The introduction of bending potentials acting on the angle between subsequent 
bonds, monomer dependent bond lengths $l_{\alpha \beta}$ 
(for bonds connecting monomers $\alpha$ and $\beta$), or
variable bond lengths with bond length potentials, etc.,
is straightforward.

In all the examples quoted so far, the chain statistics is determined
by local relations between neighbor monomers along the chain, and is
thus that of random walks. As we shall see in section 2.2, 
solving such a model in self-consistent field approximation amounts to 
solving a diffusion equation self-consistently, which can usually be
done at relatively moderate expense. This makes this kind of approach
particularly attractive. However, the self-consistent field method
is not restricted to locally defined weight functions. Szleifer and 
coworkers \cite{SZLEIFER1,SZLEIFERR} have recently devised a procedure which 
allows to perform self-consistent field calculations for chains with 
arbitrary chain statistics, {\em e.g.}, self avoiding walk statistics etc.
(see section 2.2).

In the following, we shall use the notation
\mbox{$\int\! \widehat{\cal D}_j\{\vec{R}(\cdot)\} 
= \int\! {\cal D}\{\vec{R}(\cdot)\} {\cal P}_j\{\vec{R}(\cdot)\}$}
for the weighted sum over all chain conformations, which is
a path integral in the case of continuous curves,
and a summation over discrete walks otherwise.
The partition function of the system in the canonical ensemble is 
then given by
\begin{equation}
\label{zcan0}
{\cal Z}_C =
\prod_j \Big\{ \frac{1}{n_j!} \prod_{i_j=1}^{n_j}
\int \! \widehat{\cal D}_j\{\vec{R}_{i_j}(\cdot)\} \Big\} \;
\exp[-\beta {\cal V} \{ \widehat{\rho}_{\alpha} \}],
\end{equation}
where $\beta = 1/k_B T$ is the Boltzmann factor, and the monomer interaction
functional ${\cal V} \{ \widehat{\rho}_{\alpha} \}$ has been
introduced earlier. In order to proceed, it is useful\cite{NOOLANDI1}
to insert functional integrals over delta functions 
\begin{equation}
\label{delta}
\mbox{\bf 1} = \int \! {\cal D} \{\rho_{\alpha} \} \:
\delta (\rho_{\alpha}-\widehat{\rho}_{\alpha})
=
\int \! {\cal D}  \{ \rho_{\alpha}\} 
\int_{i \infty} \!\!\!\!\!
{\cal D} \{\omega_{\alpha}\} \:
e^{ \: \int \!\! d \vec{r} \: \omega_{\alpha}(\vec{r}) \:
(\rho_{\alpha}(\vec{r}) - \widehat{\rho}_{\alpha}(\vec{r}))},
\end{equation}
which allow to replace the density operators 
$\widehat{\rho}_{\alpha}(\vec{r})$ in ${\cal V} \{ \widehat{\rho}_{\alpha} \}$ 
by density functions $\rho_{\alpha}(\vec{r})$. The subscript $\int_{i \infty}$
indicates that the limits of integration of the auxiliary fields
$\omega_{\alpha}(\vec{r})$ are $-i\infty$ to $i \infty$.
Eqn (\ref{zcan0}) can then be rewritten exactly as
\begin{equation}
\label{zcan1}
{\cal Z}_C  = 
\Big\{ \prod_{\alpha} \int \!{\cal D}  \{ \rho_{\alpha}\} 
\int_{i \infty} \!\!\!\!\! {\cal D} \{\omega_{\alpha}\} \Big\}  \:
 \exp \Big[ - \beta {\cal F}_C\{\rho_{\alpha}, \omega_{\alpha} \} \Big]
\end{equation}
with the canonical free energy functional
\begin{equation}
\label{fcan}
\beta {\cal F}_C\{\rho_{\alpha}, \omega_{\alpha} \} 
=
\beta {\cal V}\{\rho_{\alpha} \} 
- \sum_{\alpha} \int \! d \vec{r} \: 
\omega_{\alpha}(\vec{r}) \rho_{\alpha} (\vec{r})
- \sum_j n_j \ln({\cal Q}_j\{ \omega_{\alpha} \} / n_j) .
\end{equation}
The functional ${\cal Q}_j \{ \omega_{\alpha} \}$ is the partition 
function of a single chain moving in the (imaginary) external fields 
$\omega_{\alpha}(\vec{r})$,
\begin{equation}
\label{zsingle}
{\cal Q}_j \{ \omega_{\alpha} \} = 
\int \widehat{\cal D}_j\{\vec{R}(\cdot)\} \;
\exp \Big[- \sum_{\alpha} 
\int_0^{N_j} \! \! ds \: \omega_{\alpha}(\vec{R}(s)) \: \gamma_{\alpha,j}(s) 
\Big].
\end{equation}

The main step of the self-consistent field approach consists in
performing a saddle point integration of the integral (\ref{zcan1}) with
respect to $\omega_{\alpha}(\vec{r}) $: the path integral is replaced 
by the value of the integrand at the corresponding saddle function 
$\omega_{\alpha}(\vec{r})$ in the complex plane. Extremization of the exponent 
in (\ref{zcan1}) yields the equations
\begin{equation}
\label{rcan}
\rho_{\alpha} (\vec{r}) 
= - \sum_j n_j \:
\frac{\delta \ln({\cal Q}_j\{ \omega_{\alpha} \})  
}{\delta \omega_{\alpha}(\vec{r})} 
=
\big\langle \widehat{\rho}_{\alpha}(\vec{r}) \big\rangle_{C} \quad .
\end{equation}
Brackets $\langle \cdot \rangle_{C}$ denote statistical averages taken in a 
canonical ensemble of $n_j$ noninteracting chains of type $j$ subject to the 
external fields $\omega_{\alpha}(\vec{r})$. 
Note that the saddle functions $\omega_{\alpha}(\vec{r})$ which solve these 
equations are now real. The approximation thus amounts to replacing the 
exact constraint 
\mbox{$\rho_{\alpha}(\vec{r}) = \widehat{\rho}_{\alpha}(\vec{r})$}
in eqn. (\ref{delta}) by the more relaxed requirement
\mbox{$ \rho_{\alpha}(\vec{r}) = \langle \widehat{\rho}_{\alpha}(\vec{r})
\rangle_{C}$}.
The functions $\omega_{\alpha}(\vec{r})$ 
act as Lagrange parameters which enforce this condition.
The resulting partition function is 
\begin{equation}
\label{zcan2}
{\cal Z}_C  = 
\Big\{ \prod_{\alpha} \int \!{\cal D}  \{ \rho_{\alpha}\} \: \Big\} 
 \exp \Big[- \beta \; 
{\cal F}_C\{\rho_{\alpha}, \omega_{\alpha}\{\rho_{\alpha}\}\}\:\Big],
\end{equation}
where the fields $\omega_{\alpha} \{\rho_{\alpha} \}$ depend self-consistently 
on $\rho_{\alpha}$ according to eqn. (\ref{rcan}).

When looking at mixtures, it is sometimes more convenient to work in the
grand canonical ensemble\cite{SHULL1,MARK2}. The starting point is then
the grand canonical partition function
\begin{equation}
\label{zgcan0}
{\cal Z}_{GC} =
\prod_j \sum_{n_j=0}^{\infty}
\Big\{ \frac{e^{\beta \mu_j n_j}}{n_j!} \prod_{i_j=1}^{n_j}
\int \! \widehat{\cal D}_j\{\vec{R}_{i_j}(\cdot)\} \Big\} \;
\exp[-\beta {\cal V} \{ \widehat{\rho}_{\alpha} \}],
\end{equation}
where $\mu_j$ are the chemical potentials associated with molecules
of type $j$. One can now proceed the same way as above, and obtains
the analogue to (\ref{zcan1}) with the grand canonical free energy functional
\begin{equation}
\label{fgcan}
\beta {\cal F}_{GC}\{\rho_{\alpha}, \omega_{\alpha} \} 
=
\beta {\cal V}\{\rho_{\alpha} \} 
- \sum_{\alpha} \int \! d \vec{r} \: 
\omega_{\alpha}(\vec{r}) \rho_{\alpha} (\vec{r})
- \sum_j e^{\beta \mu_j} {\cal Q}_j \{ \omega_{\alpha} \}.
\end{equation}
The extremization with respect to $\omega_{\alpha}(\vec{r})$ yields the
condition
\begin{equation}
\label{rgcan}
\rho_{\alpha} (\vec{r}) 
= - \sum_j e^{\beta \mu_j}
\frac{\delta {\cal Q}_j}{\delta \omega_{\alpha}(\vec{r})} 
= \big\langle \widehat{\rho}_{\alpha}(\vec{r}) \big\rangle_{GC} \quad .
\end{equation}
Hence the grand canonical partition function in self-consistent
field approximation reads
\begin{equation}
\label{zgcan2}
{\cal Z}_{GC}  = 
\Big\{ \prod_{\alpha} \int \!{\cal D}  \{ \rho_{\alpha}\} \: \Big\} 
 \exp \Big[- \beta \; 
{\cal F}_{GC}\{\rho_{\alpha}, \omega_{\alpha}\{\rho_{\alpha}\}\}\:\Big].
\end{equation}
with $\omega_{\alpha}\{ \rho_{\alpha} \}$ defined {\em via} eqn. (\ref{rgcan}).

At this point, concentration fluctuations are still included in the
partition function. Usually, self-consistent field approximations take one 
more step and perform a second saddle point integration of (\ref{zcan2}) 
or (\ref{zgcan2}) with respect to $\rho_{\alpha}(\vec{r})$. 
The free energy is then approximated by
\begin{equation} 
\label{fscft}
F = - k_B T \ln {\cal Z} \approx F^{SCF} = 
\min_{\rho_{\alpha}(\vec{r})} 
{\cal F}\{\rho_{\alpha}, \omega_{\alpha}\{\rho_{\alpha}\}\},
\end{equation}
with the minimization equations 
\begin{equation}
\label{ww}
\omega_{\alpha}(\vec{r}) = 
\frac{\delta  \; \beta {\cal V}}{\delta \rho_{\alpha}(\vec{r})} 
\end{equation}
in both the canonical and the grand canonical ensemble. This is a very
intuitive, typical mean field result: The auxiliary fields 
$\omega_{\alpha}(\vec{r})$ which drive the monomer densities 
$\rho_{\alpha}(\vec{r})$ are identified with the local excess free energy
needed to add one monomer $\alpha$ to the melt at the position $\vec{r}$,
less the contribution of the translational entropy of polymers.
Since the fields $\omega_{\alpha}$ are functions of $\rho_{\alpha}$ 
by means of eqn. (\ref{rcan}) or (\ref{rgcan}), eqn. (\ref{ww}) effectively 
determines the mean field concentration profiles 
$\rho^{SCF}_{\alpha}(\vec{r})$.

Hence two approximations enter the self-consistent field theory:
The first leading to eqn. (\ref{zcan2}) or eqn. (\ref{zgcan2}), respectively, 
and the second leading to eqn. (\ref{fscft}). We have taken the care to
separate these two steps for the following reasons.  First, Shi, Noolandi
and Desai \cite{SHI} have recently suggested a way how to 
systematically improve on the second step and take Gaussian or even higher 
order concentration fluctuations into account (see section 2.4). This 
allows for a stability analysis of the mean field solution\cite{LARADJI}, and
for the construction of kinetic paths connecting different mean
field solutions\cite{MARK3}.  Second, the two steps differ qualitatively 
from each other. Technically, the saddle functional approximates integrals
of $\omega_{\alpha}$ on the imaginary axis, and real integrals of
$\rho_{\alpha}$. A stable saddle point is thus a {\em minimum} with
respect to $\rho_{\alpha}$ and a {\em maximum} with respect to
$\omega_{\alpha}$. The first step
motivates the introduction of a coarse-grained density functional
${\cal F}\{\rho_{\alpha}, \omega_{\alpha}\{\rho_{\alpha}\}\}$, which
is then minimized in step two as in usual mean field approaches.

The introduction of orientation dependent monomer interactions into
the formalism is straightforward. The monomer density operator (\ref{rho})
is simply replaced by 
\begin{equation}
\label{rhou}
\widehat{\rho}^j_{\alpha}(\vec{r},\vec{u})  =
\frac{1}{n_j} \sum_{i_j=1}^{n_j} \int_0^{N_j} ds \: 
\delta(\vec{r} - \vec{R}_{i_j}(s) ) \:
\delta(\vec{u} - \vec{U}_{i_j}(s) )
\: \gamma_{\alpha,j}(s) ,
\end{equation}
where $\vec{U}$ is the orientation of monomer $i_j$. The fields
$\omega_{\alpha}$ become orientation dependent and meet ({\em cf.} (\ref{ww}))
\mbox{$
\omega_{\alpha}(\vec{r},\vec{u}) = 
{\delta  \; \beta {\cal V}}/{\delta \rho_{\alpha}(\vec{r},\vec{u})}
$}.

Finally in this section, we sketch Helfand's original formulation of the 
self-consistent field theory\cite{HELFAND2}, which is particularly suited 
for the study of polymer interfaces and surfaces. Consider an interface 
between two entirely segregated homopolymer phases $j=A$ and $B$, which 
occupy each a partial volume $V_j=n_j N_j/\rho_j^*$. The two pure systems 
are characterized by their bulk densities $\rho_j^*$ and the associated fields 
$\omega_j^* = d \nu_j^*/d \rho_j^*$, where $\nu_j^*$ is the bulk
density of the interaction energy ${\cal V}$.
Let further the probability distribution ${\cal P}_j \{ \vec{R}(\cdot)\}$
be normalized such that $\int \widehat{\cal D}_j \{ \vec{R}(\cdot)\} = V_j$.
The free energy density of the pure systems is then given by
$\beta f_j^* = \beta \nu_j^* + \rho_j^*/N_j \ln (\rho_j^*/N_j)$.  
When looking at the interface between the two phases, it is now convenient 
to work with shifted monomer fields
\begin{equation}
\label{whelf}
\tilde{\omega}_j (\vec{r})= \omega_j (\vec{r}) - \omega_j^*, 
\end{equation}
which vanish in the pure system $j$. They can be interpreted as the 
excess work needed to bring a monomer of type $j$ from the pure system 
into the interfacial region. The underlying shifted interaction potential
($ \tilde{\omega}_j (\vec{r})= 
{\delta \tilde{\cal V}}/{\delta \rho_j (\vec{r})} $)
has the form
\begin{equation}
\label{vhelf}
\beta \tilde{\cal V} 
= \beta {\cal V} 
- \sum_j \omega_j^* \int \!\! d\vec{r} \: \rho_j(\vec{r})  - \mbox{const.},
\end{equation}
where ``const.''
$ = \sum_j V_j \; \big\{ \: \beta \nu_j^* - \omega_j^* \rho_j^* \: \big\}$
subtracts the bulk contribution of the pure phases. 
The shifted fields $\tilde{\omega}_j$ define a renormalized single chain 
partition function 
\begin{equation}
\tilde{\cal Q}_j =
\int \widehat{\cal D}_j \{ \vec{R}(\cdot) \} 
\exp\big[- \int_0^{N_j} \!\!\!\! ds \: \tilde{\omega}_j(\vec{R}(s)) \big]
= {\cal Q}_j \: \exp[-\omega_j^* N_j],
\end{equation}
Note that the actual value of $\tilde{\cal Q}_j$ is 
$\tilde{\cal Q}_j = n_j N_j/\rho_j^*$ 
Hence the equation for the density profiles  can
be written in the simple form
\begin{equation}
\rho_j(\vec{r}) = - \frac{\rho_j^*}{N_j} \: 
\frac{\delta \tilde{\cal Q}_j}{\delta \tilde{\omega}_j(\vec{r})},
\end{equation}
and the excess free energy of the interface is
\begin{equation}
\label{fexc}
\beta F_{exc} = \beta \tilde{\cal V} \{\rho_j\}
- \sum_j \int \! d \vec{r} \: \tilde{\omega}_j(\vec{r}) \: \rho_j(\vec{r}).
\end{equation}
The formalism can also be applied if the polymers $A$ and $B$ are not fully 
segregated, {\em i.e.}, the bulk phases contain both $A$ and $B$.
In that case, the interfacial free energy (\ref{fexc}) has to be corrected 
for additional bulk contributions. Helfand {\em et al} have approximated 
the excess interaction potential $\tilde{\cal V}$ by the simple 
form\cite{HELFAND1,HELFAND2}
\begin{equation}
\label{fhelf}
\beta \tilde{\cal V} =
\int d\vec{r} \;
\big\{ \chi \frac{\rho_A \rho_B}{\sqrt{\rho_A^* \rho_B^*}}
+ \frac{1}{2 \kappa k_B T}
( \frac{\rho_A}{\rho_A^*} + \frac{\rho_B}{\rho_B^*} -1)^2 \; \big\},
\end{equation}
where the Flory-Huggins parameter $\chi$ describes for the incompatibility
of monomers $A$ and $B$, and $\kappa$ is the compressibility of the melt.
In the limit of infinitely long polymers and zero compressibility, the
theory could be solved analytically by Helfand and Tagami
(SSL: strong segregation limit\cite{HELFAND1}). For a symmetric mixture
($b_A=b_B=b, \rho_A^*=\rho_B^*=\rho$), they obtained the interfacial tension 
$\beta \sigma_{SSL} = \sqrt{\chi/6} \: \rho b $ and the interfacial width
$w_{SSL} = b/\sqrt{6 \chi}$. 

\subsection{Implementation}

\noindent

Eqns. (\ref{rcan}) or (\ref{rgcan}) and (\ref{ww}) define a complete cycle
of self-consistent equations, which can be solved numerically by suitable 
iteration methods. The main task in each iteration step consists in evaluating 
the partition function ${\cal Q}_j\{\omega_{\alpha}\}$ of a single chain $j$
in the external fields $\omega_{\alpha}(\vec{r})$ (\ref{zsingle}) 
and the derivatives of ${\cal Q}_j\{\omega_{\alpha}\}$. 
If the polymers have random walk statistics ${\cal P}_j\{\vec{R}(\cdot)\}$
as in (\ref{pgauss}), (\ref{pworm}) or (\ref{pjoint}), it is useful to
define the propagators
\begin{eqnarray}
G_j(\vec{r},t; \vec{r}{}',t')  & = & 
\int \widehat{\cal D}_j\{\vec{R}(\cdot)\} 
\exp \Big[- \sum_{\alpha} 
\int_t^{t'} \! \! ds \: \omega_{\alpha}(\vec{R}(s)) \: \gamma_{\alpha,j}(s) 
\Big] \;
\nonumber  \\
&&
\times \;
\delta(\vec{r}-\vec{R}(t)) \;
\delta(\vec{r}{}'-\vec{R}(t')) ,
\end{eqnarray}
or, in case monomer orientations are important,
\begin{eqnarray}
G_j(\vec{r},\vec{u},t; \vec{r}{}',\vec{u}{}',t')  & = & 
\int \widehat{\cal D}_j\{\vec{R}(\cdot)\} 
\exp \Big[- \sum_{\alpha} 
\int_t^{t'} \! \! ds \: \omega_{\alpha}(\vec{R}(s)) \: \gamma_{\alpha,j}(s) 
\Big] \;
\nonumber  \\
\lefteqn{\hspace{-1cm} \times \;
\delta(\vec{r}-\vec{R}(t)) \;
\delta(\vec{r}{}'-\vec{R}(t')) \;
\delta(\vec{u}-\vec{U}(t)) \;
\delta(\vec{u}{}'-\vec{U}(t')). }
\end{eqnarray}
Since they carry no memory along the chain, these propagators
satisfy modified diffusion equations with the initial condition
\mbox{$G_j(\vec{r},t; \vec{r}{}',t) =  \delta(\vec{r}-\vec{r}{}')$} or
\mbox{$G_j(\vec{r},\vec{u},t; \vec{r}{}',\vec{u}{}',t') =  
\delta(\vec{r}-\vec{r}{}') \; \delta(\vec{u}-\vec{u}{}')$}.
Specifically, the diffusion equation for Gaussian chains (\ref{pgauss})
reads \cite{HELFAND1}
\begin{equation}
\Big[
\frac{\partial}{\partial t'} +
\sum_{\alpha} \gamma_{\alpha,j}(t')
\big[
- \frac{b_{\alpha}^2}{6} \nabla^2_{\vec{r}{}'}
+ \omega_{\alpha}(\vec{r}{}') 
\big]
\Big] \:
G_j(\vec{r},t; \vec{r}{}',t')  =  0;
\end{equation}
for wormlike chains (\ref{pworm})\cite{MORSE}, 
\begin{equation}
\Big[
\frac{\partial}{\partial t'} +
\sum_{\alpha} \gamma_{\alpha,j}(t')
\big[
a_{\alpha} \vec{u}{}' \nabla_{\vec{r}{}'}
- \frac{1}{2 \eta_{\alpha}} \nabla^2_{\vec{u}{}'}
+ \omega_{\alpha}(\vec{r}{}') 
\big]
\Big] \:
G_j(\vec{r},\vec{u},t; \vec{r}{}',\vec{u}{}',t')  =  0,
\end{equation}
and for freely jointed chains with bond length $l$ (\ref{pjoint}), 
\begin{equation}
G_j(\vec{r},t; \vec{r}{}',t'+1)  =  
\exp[-\omega_{\alpha}(\vec{r}{}')] \:
\frac{1}{4 \pi} \int d \vec{u} \: 
G_j(\vec{r},t; \vec{r}{}'- l\: \vec{u},t'),
\end{equation}
where $|\vec{u}|=1$ as before, and $\int \! d \vec{u}$ denotes integration 
over the full solid angle. Once the propagators have been determined, one 
easily calculates the partition function 
\mbox{${\cal Q}_j=\int \!\! d \vec{r}\:d \vec{r}{}'
\: G_j(\vec{r},0;\vec{r}{}',N_j)$}
and its derivatives
\begin{eqnarray}
\frac{\delta {\cal Q}_j}{\delta \omega_{\alpha}(\vec{r}_0)}
&=& \int \!\ d \vec{r} \: d \vec{r}{}' 
\int_0^{N_j} \!\!\! ds \: \gamma_{\alpha,j}(s) \:
G_j(\vec{r},0;\vec{r}_0,s) \: G_j(\vec{r}_0,s;\vec{r}{}',N_j) 
\nonumber \\
\frac{\delta^2 {\cal Q}_j}
{\delta \omega_{\alpha}(\vec{r}_0) \delta \omega_{\beta}(\vec{r}_1)}
&=& 2 \int \! d \vec{r} \: d \vec{r}{}' 
\int_0^{N_j} \!\!\! ds \: \gamma_{\alpha,j}(s) \:
\int_0^s \!\! ds' \: \gamma_{\beta,j}(s') \: \\
&& \qquad G_j(\vec{r},0;\vec{r}_0,s') \; G_j(\vec{r}_0,s';\vec{r}_1,s) \;
G_j(\vec{r}_1,s;\vec{r}{}',N_j) \nonumber
\end{eqnarray}
etc. In fact, only the first derivatives enter the self-consistent field 
equations. It is thus often less time consuming to 
calculate directly the ``end segment distributions'' 
$q_j(\vec{r},t)=\int \!\! d \vec{r}{}' G_j(\vec{r}{}',0; \vec{r},t)$ and 
$\bar{q}_j(\vec{r},t)=\int \!\! d \vec{r}{}' G_j(\vec{r},t; \vec{r}{}',N_j)$,
which solve the same diffusion equation as $G$ with boundary conditions
\mbox{$q_j(\vec{r},0)=1$} and $\bar{q}_j(\vec{r},N_j)=1$. The end segment 
distribution functions can be used to calculate density profiles, chain end 
distributions etc. The full propagators $G_j(\vec{r},t;\vec{r}{}',t')$ 
are needed when looking at the effect of fluctuations 
(see section 2.4), or at chain correlation functions such as distributions 
of end-to-end vectors.

Diffusion equations can usually be solved without too much computational
effort. The computer time needed for one iteration step
only scales linearly with the chain length $N$, hence one can handle
relatively long chains. Physically, treating the chains as random 
walks is well motivated in dense melts of long polymers, since 
the excluded volume interactions between monomers of the same chain are 
screened by the presence of other chains\cite{DEGENNESB}.

However, this is not true for dilute polymer chains, and on length 
scales smaller than the screening length\cite{DEGENNESB}. 
The random walk approximation thus becomes questionable when looking
at polymers in good solvent conditions, or at relatively short 
macromolecules. In that case, it is more appropriate to choose a weight
distribution ${\cal P}_j$ which accounts also
for long range correlations within a chain. One then has to consider whole 
chains in the external field of the other chains (single chain mean field). 

Szleifer and Coworkers have recently suggested an enumeration procedure for 
the evaluation of chain partition functions \cite{SZLEIFER1,SZLEIFERR}. 
They approximate the 
path integral $\int \widehat{\cal D}_j \{\vec{R}(\cdot)\}$ by the sum 
over a representative sample of chain configurations $\vec{R}_{i_j}(\cdot)$, 
which are distributed according to the weight function 
${\cal P}_j\{\vec{R}(\cdot)\}$. 
The sample can be generated by Monte Carlo simulations, taken from 
experiments etc. This approach is conceptually extremely simple and can be 
applied to arbitrary weight distributions ${\cal P}_j\{\vec{R} (\cdot) \}$. 
For example, dilute polymers in good solvent are well described by self 
avoiding walks, and one can draw the sample from single chain 
simulations\cite{SZLEIFER1}.
Semidilute polymers have self avoiding walk statistics on short length 
scales below the screening length, and random walk statistics on larger 
length scales. The best description on all length scales is obtained when
the sample is generated in Monte Carlo simulations of a melt\cite{MARCUS1}.
Unfortunately, the computational effort required to reliably solve the 
self-consistent equations grows rapidly with the chain length, since the
sizes of the sample have to be made very large. 
In dense melt of long polymers, random 
walk models for chains usually give reasonably good results at much less 
expense. 

Nevertheless, the method is very versatile and has much potential for 
further refinement. For example, reweighting schemes can be devised in 
analogy to the reweighting techniques in Monte Carlo simulations\cite{BERG}, 
which are useful
if the  actual distribution of chain conformations differs very much
from the distribution ${\cal P}_j\{\vec{R} (\cdot) \}$. One then draws the 
sample of chain conformations according to a modified distribution
${\cal P'}_j\{\vec{R}(\cdot) \}$, and corrects for this in the sum
over conformations.
\begin{displaymath}
\textstyle
\int \widehat{\cal D}_j \{\vec{R}(\cdot)\} \cdots
\longrightarrow 
\sum_{\vec{R}_{i_j}} 
{\cal P}_j\{\vec{R}_{i_j}(\cdot) \}/{\cal P'}_j\{\vec{R}_{i_j}(\cdot) \} 
\; \cdots
\end{displaymath}
Weinhold {\em et al} have introduced a coupling between ${\cal P}_j$ and the 
number of interchain contacts, and could qualitatively reproduce the shrinking 
of chains in an athermal melt with increasing monomer density\cite{WEINHOLD}. 
Within a similar approach, M\"uller calculates the statistical mechanics of 
clusters of polymers embedded in a self-consistent field, in order to study 
the conformations of a single $A$ chain in an unfavorable $B$ 
environment\cite{MARCUS2}. His results compare well with 
Monte Carlo simulations.

With these remarks, we end the discussion of the Szleifer method. The
reader is referred to \cite{SZLEIFERR} for a detailed review of the approach 
and its applications. We close this section with a few technical comments
on the solution of the mean field equations. 

The careful choice of the suitable basis is crucial for the success of a
self-consistent field calculation. Matsen {\em et al} have implemented with 
great success a method to study periodically ordered mesophases, where 
spatially-dependent functions are expanded in orthonormal basis functions 
with the symmetry of the phase being considered\cite{MARK4,MARK5,MARKR}. 
If monomer interactions come into play, it is usually beneficial to expand
functions of the orientation $\vec{u}$ in spherical harmonics\cite{FS1}.

The mean field equations are solved iteratively, {\em e.g.}, using
the fields $\omega_{\alpha}(\vec{r})$ as iteration variables. The most
popular iteration procedure is the Newton Raphson method\cite{recipes}. 
One usually obtains a solution of reasonably high accuracy
within less than 10 iteration steps. However, each step involves the 
determination of a full Hesse matrix of derivatives, {\em i.e.},
the number of function evaluations grows quadratically with the number
of degrees of freedom. Relaxation procedures take more iteration steps,
but may nevertheless require fewer function evaluations in the end.
The present author has had good experience with a variant of a method
originally suggested by Ng\cite{NG,FS2}. Another efficient procedure is the
multidimensional secant method of Broyden\cite{numerical2,PHILIP1}.


\subsection{Monomer interactions and compressibility}

\noindent

We now turn to the discussion of the functional ${\cal V}\{\rho_{\alpha}\}$, 
which defines together with ${\cal P}_j\{ \vec{R}(\cdot)\}$ the actual model. 
It describes the free energy of a system of interacting monomers without 
translational entropy. Naively, one could be tempted to identify it with the 
sum of direct interactions $W_{\alpha \beta}(\vec{r})$ between monomers,
$\frac{1}{2}\int \!\! d \vec{r} \: d \vec{r}{}' \sum_{\alpha \beta} 
W_{\alpha \beta}(\vec{r}-\vec{r}{}')\rho_{\alpha}(\vec{r})
\rho_{\beta}(\vec{r}{}')$.
However, this approximation is 
poor in dense fluids, where indirect interactions and multiplet correlations 
are important. In most applications of the self-consistent field theory,
the quantities of interest are composition inhomogeneities, whereas the total 
density varies comparatively little throughout the system. Furthermore, the
relative repulsion of monomers of different type is usually vanishingly
small compared to the total free energy in the melt.
It is thus propitious to separate ${\cal V}$ into an ``equation of state''
contribution ${\cal V}_0$, which accounts for density fluctuations and 
compressibility effects, and an interaction part 
${\cal V}_{\mbox{\tiny inter}}$, which  incorporates the incompatibility 
of unlike monomers,
\begin{equation}
{\cal V}\{\rho_{\alpha}\} = {\cal V}_0 \{ \rho_{\alpha} \} 
+ {\cal V}_{\mbox{\tiny inter}} \{ \rho_{\alpha} \}.
\end{equation} 

We shall discuss both parts in turn, and begin with ${\cal V}_0$.
If the melt is nearly incompressible and the density fluctuations are 
small, one may use Helfand's quadratic approximation\cite{HELFAND1,HELFAND2}
(see eqn. (\ref{fhelf}))
\begin{equation}
\label{vparts}
{\cal V}_0 \{\rho_{\alpha} \} =
\frac{1}{2 \kappa}
\int d\vec{r} \;
(\; 1 - \sum_{\alpha} \rho_{\alpha}\: v_{\alpha} \; )^2 ,
\end{equation}
where $v_{\alpha}$ is the specific volume of an $\alpha$ monomer. 

More rigorously, total incompressibility is often postulated.
\begin{equation}
\label{incomp}
{\sum_{\alpha} \rho_{\alpha}(\vec{r}) \: v_{\alpha} \equiv 1}
\qquad \mbox{everywhere.}
\end{equation}
This is most commonly achieved by introducing an additional Lagrange 
parameter field $\xi(\vec{r})$, which couples to the incompressibility
constraint (\ref{incomp})\cite{NOOLANDI1,SHULL1,MARK4}.
The free energy functional (\ref{fcan}) or (\ref{fgcan}) is then
replaced by
\begin{eqnarray}
\label{finc}
\beta {\cal F}
&=&
\textstyle
\beta {\cal V}_{\mbox{\tiny inter}}\{\rho_{\alpha} \} 
- \sum_{\alpha} \int \! d \vec{r} \; 
\omega_{\alpha}(\vec{r}) \rho_{\alpha} (\vec{r})
- \int \! d \vec{r} \; \xi(\vec{r}) \;
\big[ \sum_{\alpha} \rho_{\alpha} (\vec{r}) \: v_{\alpha} - 1 \big]
\nonumber\\
 &&- \:\left\{ 
\begin{array}{ll}
\sum_j n_j \ln({\cal Q}_j\{ \omega_{\alpha} \} / n_j) 
& \mbox{\quad (canonical ensemble)} \\
\sum_j e^{\beta \mu_j} {\cal Q}_j \{ \omega_{\alpha} \}.
& \mbox{\quad (grand canonical ensemble),}
\end{array}
\right.
\end{eqnarray}
and the minimization equation (\ref{ww}) turns into
\begin{equation}
\omega_{\alpha}(\vec{r}) = 
\frac{\delta  \; \beta {\cal V}_{\mbox{\tiny inter}}}
{\delta \rho_{\alpha}(\vec{r})} - \xi(\vec{r}) \: v_{\alpha}.
\end{equation}
The incompressibility constraint (\ref{incomp}) determines the field 
$\xi(\vec{r})$ unambiguously, except for a constant $\xi_0$. 
It is often chosen such that $\int \! d\vec{r}\: \xi(\vec{r}) = 0$.
Note that the choice of $\xi_0$ determines the offset of the
chemical potentials $\mu_j$ in the grand canonical ensemble.

Within this framework of an incompressible theory, total density fluctuations 
is often reintroduced by means of a hypothetical noninteracting 
``solvent'' $S$, which fills the space between the 
molecules\cite{NOOLANDI1,SZLEIFERR,THEODOROU,HARIHARAN}. 
The incompressibility condition (\ref{incomp}) then reads
\mbox{$\sum_{\alpha} \rho_{\alpha}(\vec{r}) \: v_{\alpha} 
+ \rho_s(\vec{r}) v_s \equiv 1$}, and the excess potential of the solvent
is simply given by \mbox{$\omega_s(\vec{r}) = \xi(\vec{r}) \: v_s$}. 
According to eqn. (\ref{rcan}) or (\ref{rgcan}), the density of the solvent is
thus given by \mbox{$\rho_s(\vec{r}) = z \: \exp[-\xi(\vec{r}) \: v_s]$},
with \mbox{$z=n_s/\int \! \! d \vec{r} \exp[-\xi(\vec{r}) v_s]$} in
the canonical case and $z = \exp[\beta \mu_s]$ in the grand canonical
case. This defines $\xi(\vec{r})$ as a function of the monomer densities 
$\rho_{\alpha}(\vec{r})$. Choosing $\xi_0$ such that \mbox{$\ln(z v_s)= 0$}, 
one obtains the fields
\begin{equation}
\omega_{\alpha}(\vec{r}) = 
\frac{\delta  \; \beta {\cal V}_{\mbox{\tiny inter}}}
{\delta \rho_{\alpha}(\vec{r})} 
- \frac{v_{\alpha}}{v_s} \; 
\textstyle
\ln(1-\sum_{\alpha} \rho_{\alpha}(\vec{r}) \: v_{\alpha} ).
\end{equation}
The approach amounts to using an equation of state potential of the form
\begin{equation}
\label{sl}
\beta {\cal V}_0 \{\rho_{\alpha} \} =
\frac{1}{v_s}
\int \! d\vec{r} \;
(\; 1 - \eta(\vec{r}) \;)
\ln \;(\; 1 - \eta(\vec{r}) \; )
\displaystyle
+ \frac{1}{v_s} \int \! d \vec{r} \: \eta(\vec{r}),
\end{equation}
where the packing fraction
$\eta(\vec{r})= \sum_{\alpha} \rho_{\alpha}(\vec{r})\: v_{\alpha}$
is the local volume fraction occupied by monomers. The last term ensures
that the free energy per monomer vanishes in the limit $\eta \to 0$.
Eqn. (\ref{sl}) is a local version of the familiar compressible Flory-Huggins 
or Sanchez-Lacombe theory\cite{SANCHEZ1}. It provides an intuitive and 
straightforward treatment of compressibility effects in polymer blends. 
Unfortunately, the prediction for the equation of state of the melt can be
rather poor\cite{DEUTSCH}, due to the total neglect of details of the 
monomer structure and local monomer correlations.

An alternative approach suggested in Ref. \cite{FS3} conceives ${\cal V}_0$ 
as the density functional of a suitable reference system of identical 
monomers with no translational entropy.
In local density approximation, ${\cal V}_0$ can be derived 
from the equation of state $\Pi(\rho)$ ($\Pi$ is the pressure),
\begin{equation}
\beta{\cal V}_0 = \int \! d \vec{r} \; \rho(\vec{r}) \; 
f[\rho(\vec{r})] 
\qquad \mbox{with} \qquad
f(\rho) = \int_0^{\rho} \!\! dx \; \frac{\Pi(x)}{x^2}.
\end{equation}
The function $f(\rho)$ is the local free energy per monomer. When
studying interfaces or surfaces, ${\cal V}_0$ is conveniently replaced
according to (\ref{vhelf}) by 
\begin{equation}
\beta \tilde{\cal V}_0 = \int \! d \vec{r} \; \big[ \; \rho(\vec{r}) \; 
( \; f[\rho(\vec{r})] - \omega^* )\; - \mbox{const.} \big],
\end{equation}
where $\omega^* = f(\rho^*) + \rho^* (df/d\rho)_{\rho^*}$ is the
excess free energy per monomer at the bulk density $\rho^*$,
and \mbox{const. $= \rho^*{}^2 (df/d\rho)_{\rho^*}$}. 
The theory requires as input the knowledge of the equation of state of the 
reference system, {\em i.e.}, a melt of infinitely long polymers 
built from the reference monomers. For example, one can take the
Flory-Huggins form\cite{FLORY1}
\begin{equation}
\Pi(\rho) \: /\rho  = - 1 - 1/\eta \: \ln (1-\eta),
\end{equation}
with the packing fraction $\eta = \rho/\rho^*$, and
recover the potential (\ref{sl}) with $v_s = 1/\rho^*$. Based on
the Carnahan-Starling equation of state for single hard spheres, Dickman 
and Hall have derived an equation of state for hard chains\cite{DICKMAN}, 
which compares well with computer simulations of hard chains, and
even of the bond fluctuation model\cite{DEUTSCH},
\begin{equation}
\Pi(\rho) \: /\rho  = C_0 \left[
\frac{1+\eta+\eta^2-\eta^3}{(1-\eta)^3}-1 \right],
\end{equation}
and leads to the free energy
\begin{equation}
f(\rho) = C_0 \; \eta\;(4-3 \eta)/(1-\eta)^2.
\end{equation}
The constant $C_0$ depends on details of the intramolecular chain
structure, and $\eta = a^3 \rho \pi/6$ is the actual volume occupied by
monomers of diameter $a$. Other forms for the equation of
state are available as well\cite{eos}. In some cases, one may wish to 
include more information on the local liquid structure in the 
functional ${\cal V}_0$, and in the corresponding monomer excess free energy
$\omega(\vec{r})$. Nath {\em et al}\cite{NATH1} suggest a modified version
of the self-consistent field theory, where the excess free energy is
given by (cf. (\ref{whelf}))
\begin{equation}
\tilde{\omega}(\vec{r}) = 
- \int \!\! d \vec{r}{}' \: c(\vec{r}-\vec{r}{}') \: 
[\; \rho(\vec{r})-\rho^* \;] + \omega^*,
\end{equation}
with the direct correlation functions $c(\vec{r})$ taken from 
P-RISM theory\cite{SCHWEIZERR}.

Next we discuss the second contribution ${\cal V}_{\mbox{\tiny inter}}$ to 
the free energy functional ${\cal V}\{\rho_{\alpha}\}$, which describes the 
monomer specific part of the interactions. Since it is small compared
to the total free energy of the system, we may assume a perturbative 
treatment. We split the pair interactions $W_{\alpha \beta}(\vec{r})$ 
between monomers $\alpha$ and $\beta$ into two parts 
$W_{\alpha \beta}(\vec{r}) = W_0(\vec{r}) + W'_{\alpha \beta}(\vec{r})$.
The potential $W_0(\vec{r})$ describes the interactions in the
reference system ${\cal V}_0$. Then we define
\begin{equation}
\label{uu}
\beta \: U_{\alpha \beta}(\vec{r}) = 
1 - \exp[-\beta \: W'_{\alpha \beta}(\vec{r})].
\end{equation}
If $W'_{\alpha \beta}(\vec{r})$ is integrable and small, 
$U_{\alpha \beta}(\vec{r})$ reduces to $W'_{\alpha \beta}(\vec{r})$.
More generally, the expression
(\ref{uu}) also allows to deal with effects of non-additive packing, 
size disparities of monomers and the like (see also Ref. \cite{MARCUSR})
In perturbation theory\cite{HANSEN},
the free energy contribution ${\cal V}_{\mbox{\tiny inter}}$ is given by
\begin{equation}
{\cal V}_{\mbox{\tiny inter}} \{ \rho_{\alpha} \}  = 
\frac{1}{2} \int \! d\vec{r} \: d \vec{r}{}' 
\sum_{\alpha \beta}  U_{\alpha \beta}(\vec{r}-\vec{r}{}') \; 
\rho_{\alpha \beta}^{(2)}(\vec{r},\vec{r}{}').
\end{equation}
Here $\rho_{\alpha \beta}^{(2)}$ is the pair density of type $\alpha$
monomers at $\vec{r}$ and type $\beta$ monomers at $\vec{r}{}'$,
which are {\em not} direct neighbors along one polymer chain. 
(The interactions between the latter contribute to the 
conformational weight functional ${\cal P}_j\{\vec{R}(\cdot)\}$).
Furthermore, the pair distribution function is approximated by
\begin{equation}
\label{gamma}
\rho_{\alpha \beta}^{(2)}(\vec{r},\vec{r}{}') =
\rho_{\alpha}(\vec{r}) \: \rho_{\beta}(\vec{r}{}') \: 
\gamma(\vec{r}-\vec{r}{}'),
\end{equation}
{\em i.e.}, the monomer pair correlation function $\gamma(\vec{r})$ is
taken to be independent of the identity of the interacting monomers, and 
of their densities\cite{FN1}.

If the interactions are short range, the profiles $\rho_{\beta}(\vec{r}{}')$ 
can be expanded around $\vec{r}$.\cite{HELFAND2}
With the definitions
\begin{eqnarray}
\chi_{\alpha \beta} &=& \beta \rho^*
\int \! d \vec{r} \: U_{\alpha \beta}(\vec{r}) \: \gamma(\vec{r}) \\
\sigma_{\alpha \beta}^{ij} &=& \beta \rho^*
\int \! d \vec{r} \: U_{\alpha \beta}(\vec{r}) \: \gamma(\vec{r}) \:
r_i \: r_j
\end{eqnarray}
etc., one obtains
\begin{equation}
\beta {\cal V}_{\mbox{\tiny inter}} \{ \rho_{\alpha} \}  = 
\frac{1}{2 \: \rho^*} \int \! d\vec{r} \: 
\sum_{\alpha \beta}  \big[ \rho_{\alpha}(\vec{r}) \: \rho_{\beta}(\vec{r}) \:
\chi_{\alpha \beta} 
- \frac{1}{2} \sum_{ij}
\frac{d \rho_{\alpha}}{d r_i} \frac{d \rho_{\alpha}}{d r_j}
\sigma_{\alpha \beta}^{ij} + \cdots \big].
\end{equation}
The indices $i$ run over cartesian coordinates $x,y,z$, and the bulk
density $\rho^*$ was introduced in order to make $\chi_{\alpha \beta}$
dimensionless. The resulting excess potentials are given by
\begin{equation}
\omega_{\alpha}(\vec{r})
= \frac{\delta {\beta \cal V}_0}{\delta \rho} 
+ \frac{1}{\rho^*}\sum_{\beta} 
\big[ \: \chi_{\alpha \beta} \: \rho_{\beta}(\vec{r})
+ \frac{1}{2} \sum_{ij} \sigma_{\alpha \beta}^{ij} 
 \frac{d^2 \rho_{\beta}}{d r_i d r_j} + \cdots \: \big].
\end{equation}
In systems with only two types of monomers $A$ and $B$, the parameters
$\chi_{\alpha \beta}$ are usually combined to one single 
``Flory-Huggins parameter'' 
\begin{equation}
\chi = \chi_{AB} - \frac{1}{2} (\chi_{AA} + \chi_{BB}).
\end{equation}
Similarly, we define $\sigma^{ij} = \sigma_{AB}^{ij} 
- \frac{1}{2} (\sigma_{AA}^{ij} + \sigma_{BB}^{ij})$.
The fields $\omega_{A,B}$ can then be rewritten as
\begin{eqnarray}
\omega_A(\vec{r}) &=&
\frac{\delta {\beta \cal V}_0}{\delta \rho} 
+ \frac{1}{\rho^*} \chi \; \rho_B 
+ \frac{1}{2 \rho^*} \sum_{ij} \sigma^{ij} \; 
\frac{d^2 \rho_B}{dr_i dr_j} 
+ h\{\rho(\vec{r})\}
\\
\omega_B(\vec{r}) &=&
\frac{\delta {\beta \cal V}_0}{\delta \rho} 
+ \frac{1}{\rho^*} \chi \; \rho_A 
+ \frac{1}{2 \rho^*} \sum_{ij} \sigma^{ij} \; 
\frac{d^2 \rho_A}{dr_i dr_j} 
- h\{\rho(\vec{r})\},
\nonumber
\end{eqnarray}
\begin{equation}
\mbox{where} \qquad
\label{hh}
h\{\rho(\vec{r})\} =
\frac{1}{4} (\chi_{AA}-\chi_{BB}) \: \rho 
+ \frac{1}{8} \sum_{ij} (\sigma_{AA}^{ij}-\sigma_{BB}^{ij}) \: 
\frac{d^2 \! \rho}{dr_i dr_j}
\end{equation}
depends only on the total density profile $\rho(\vec{r}) = \rho_A+\rho_B$,
and contributions which are identical for both components have been dropped 
except for the leading term ${\delta {\beta \cal V}_0}/{\delta \rho}$. 
Similarly, the free energy functional ${\cal V}_{\mbox{\tiny inter}}$ reads
\begin{equation}
\beta {\cal V}_{\mbox{\tiny inter}}  = 
\frac{1}{\rho^*} \int \! d\vec{r} \: \Big[ \:
\chi \; \rho_A \rho_B  
- \frac{1}{2} \sum_{ij} \sigma^{ij} 
\frac{d \rho_A}{d r_i} \frac{d \rho_B}{d r_j}
+ h\{\rho\} \; (\rho_A-\rho_B) 
+ v\{\rho\} \;
\Big]. \nonumber
\end{equation}
Here again, $v\{\rho\}$ can be neglected compared to $\beta {\cal V}_0$.

Usually, only the ``local'' contributions ($\propto \chi_{\alpha \beta}$) are 
taken into account. However, the higher order nonlocal terms become important 
in situations where the local density profiles vary strongly. We shall 
illustrate this with the example of surface segregation in a completely 
symmetric binary ($A,B$) blend of two incompatible, but otherwise identical 
homopolymers. Even if the surface is chosen neutral, {\em i.e.}, the surface 
tensions of $A$ and $B$ are equal, the minority component segregates to the 
surface, because it has less unfavorable contacts with polymers of the 
majority component there. This effect is often called a
``missing neighbor effect''. It increases with the range of the 
interactions, {\em i.e.}, it should increase with the value of $\sigma$. 

\begin{figure}

\fig{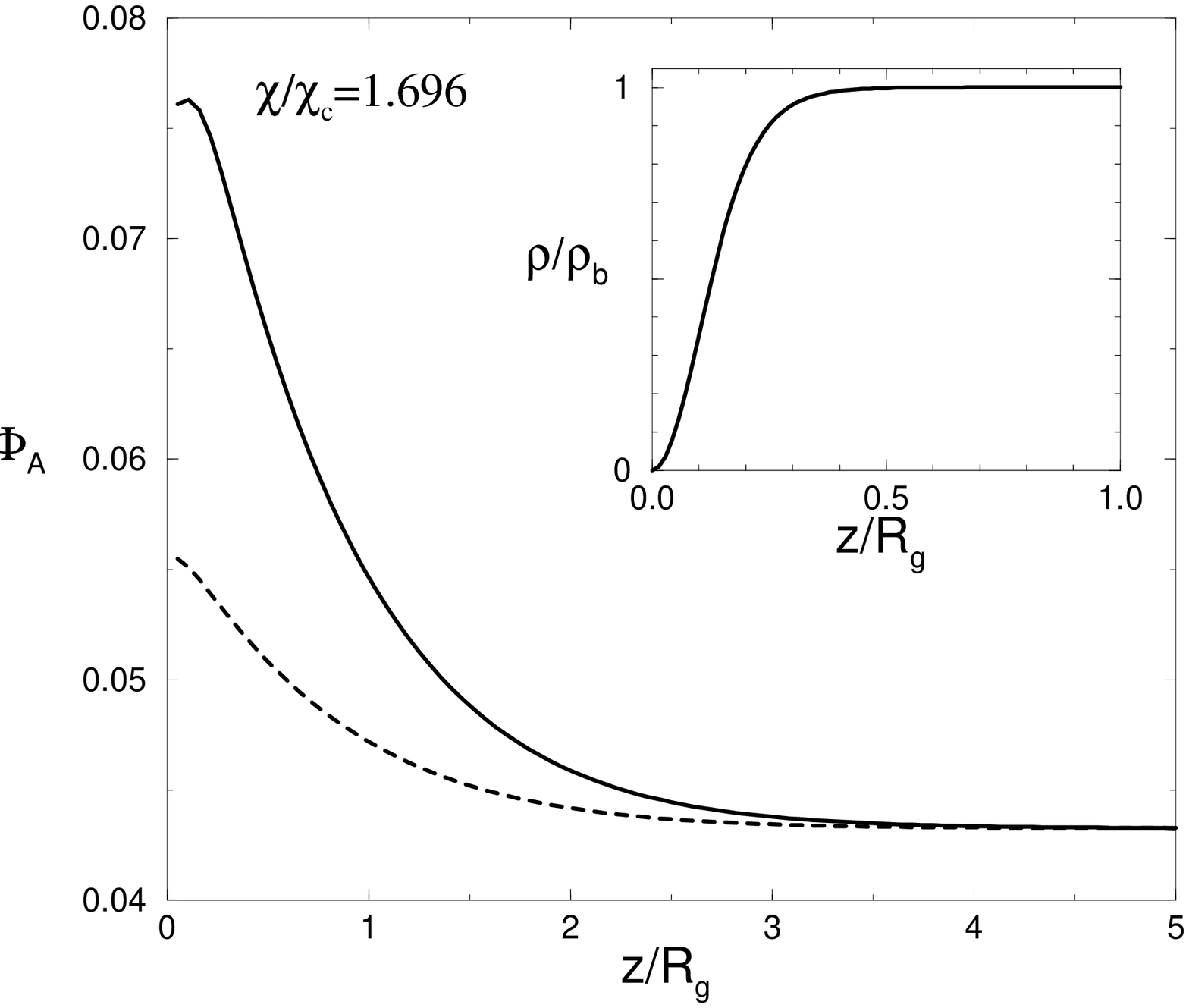}{100}{80}

\caption{\label{rho-0.02}}
Surface segregation profile of the minority component at coexistence below
the demixing transition ($\chi_c$) in a symmetric polymer mixture.
Dashed line shows the volume fraction profile for $\sigma=0$, where
the range of monomer interactions is assumed to be zero. The inset
shows the total density profile.

from Ref. \cite{FS3}

\end{figure}

In Ref. \cite{FS3} self-consistent field calculations are presented for
such a system. The parameters of the theory were adjusted to the 
bond fluctuation model, a lattice model for polymers which has been widely 
used in Monte Carlo simulations and has extremely well known bulk 
properties. It will be introduced in some more detail in Section 3.2. 
The chain conformations in the bulk\cite{PAUL}, the equation of 
state\cite{DEUTSCH,MARCUS3}, the thermodynamics of mixing\cite{MARCUS4} 
have been investigated in detail. 
Therefore, the model parameters for the self-consistent theory, {\em e.g.},
within the Gaussian chain model, are all known. Fig. \ref{rho-0.02} shows 
two segregation profiles at bulk two-phase coexistence,
one with $\sigma=0$ (dashed line) and one with $\sigma$ adapted to the bond 
fluctuation model (solid line).
(Note that the field $h\{\rho\}$ (\ref{hh}), which couples directly to the
composition fluctuations, vanishes in a perfectly symmetric mixture.)
In the limit of $\sigma=0$ or pure contact interactions, surface segregation
is almost entirely suppressed. Using the adjusted value of $\sigma$, one finds
that the volume fraction $\Phi_A=\rho_A/\rho$ of the minority phase increases 
by almost a factor of two at the surface.

\begin{figure}

\fig{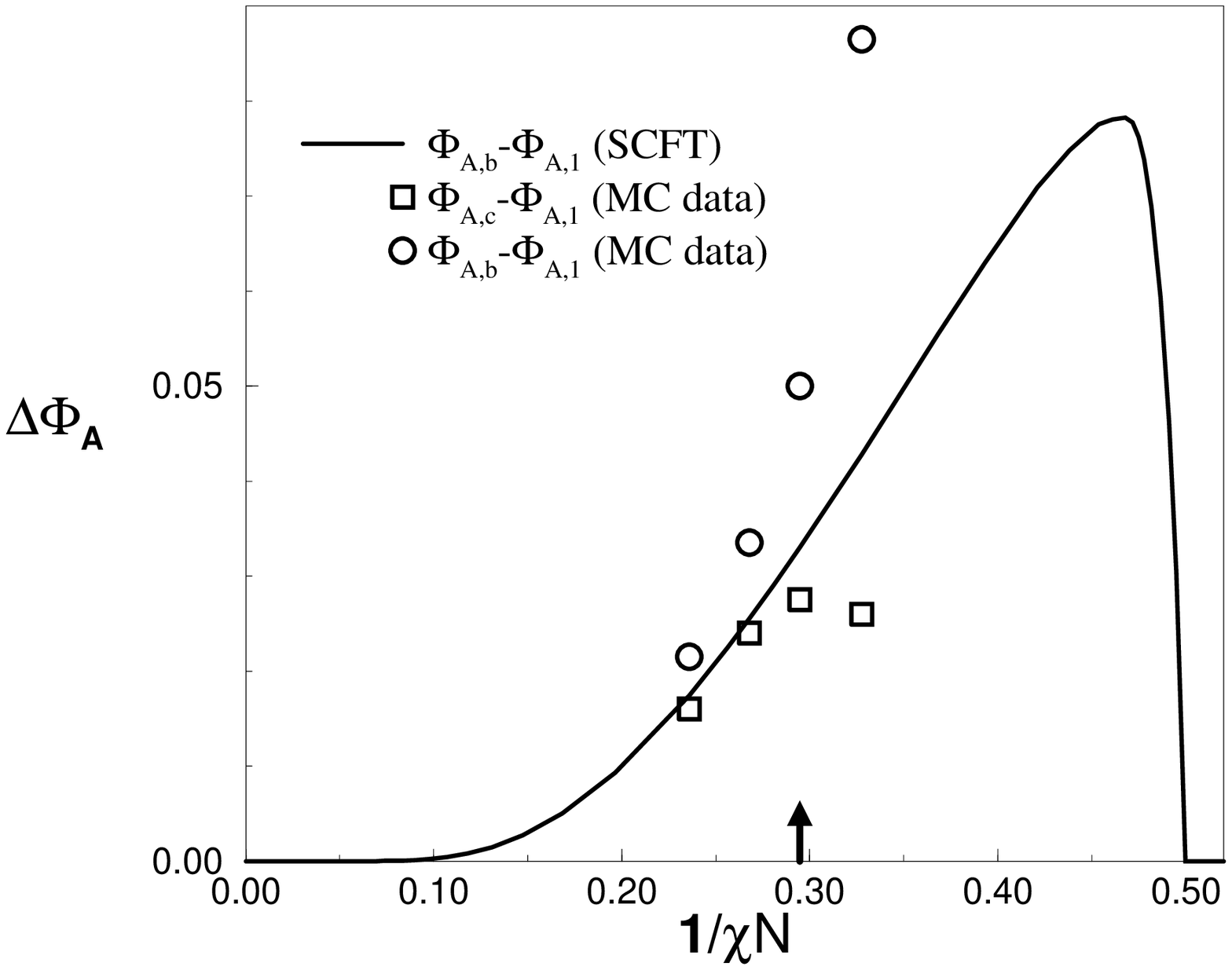}{100}{80}

\caption{\label{dpa_t}}
Difference $\Delta \Phi_A$ between the volume fraction of the minority
component at the surface and in the bulk, at two-phase coexistence,
vs. $1/\chi N$ ($N=32$). Data points show the simulation results of Rouault
{\em et al} (Ref. \cite{FS3}): Upper bound for $\Delta \Phi_A$ (circles),
and lower bound (squares). The arrow indicates the value of $1/\chi N$
which corresponds to Fig. \ref{rho-0.02}. 

from Ref. \cite{FS3}

\end{figure}

The self-consistent field results can be compared with Monte Carlo
simulations of Rouault {\em et al}\cite{YANNICK}. Fig. \ref{dpa_t} shows the 
predicted difference between the volume fractions of $A$ at the surface and
in the bulk $\Delta \Phi_A$, and corresponding Monte Carlo data.
The simulations were performed in a slab geometry with a slab thickness of 
about three times the gyration radius of the chains. In such a thin film, 
the composition variable $|\rho_A - \rho_B|$ at coexistence is reduced 
compared to the bulk system. The difference between the volume fraction 
$\Phi_A$ at the surface of the thin film, and in the bulk of the infinite 
system, thus gives an upper bound for the actual value of $\Delta \Phi_A$.
The difference between $\Phi_A$ at the surface and in the center of the
thin film gives a lower bound. Fig. \ref{dpa_t} illustrates that the 
theoretical prediction lies nicely within these two bounds. We emphasize that
the good agreement was reached without any adjustable parameters. 
However, the range of the monomer interactions has to be accounted for
correctly, {\em i.e.}, the nonlocal term $\sigma$ may not be neglected
in this situation with strong density variations.

A second {\em caveat} applies to the approximation (\ref{gamma}). In
general, the pair correlation function depends on the identity of the 
monomers, and on the local environment. For example, the number of monomer 
contacts divided by the local density typically decreases in the vicinity 
of an interface\cite{MARCUS5}, and is different for homopolymers and
copolymers\cite{ANDREAS1}. This is a higher order effect in our perturbative
approach, which is usually not taken into account in self-consistent field
theories. Wayne {\em et al} have recently found experimental
evidence that it may engender a difference of the effective $\chi$ parameter 
in homopolymer and block copolymer blends\cite{WAYNE}.

\subsection{Fluctuations}

In the limit of infinite chain length (while keeping the densities constant), 
the saddle point integrations leading to the free energy expression 
(\ref{fscft}) become exact. At any finite chain length, however, the free
energy is affected by concentration fluctuations. On the one hand, the 
concentrations always fluctuate locally about their local mean value. The 
length scale for these fluctuations is given by the bulk correlation length, 
which is usually roughly the gyration radius and diverges only very close to a
critical point. In systems with interfaces, a second qualitatively different 
type of fluctuations emerges: Since the interface breaks a continuous 
symmetry, the translational symmetry, long wavelength goldstone excitations 
come into existence which cost virtually no energy in the zero wavelength 
limit. The corresponding length scale thus diverges at all temperatures.
These fluctuations are sometimes referred to as interface ``waviness'', or 
capillary wave fluctuations.

Local concentration fluctuations can be assessed systematically by an
expansion of local concentration perturbations,
$\delta \rho_{\alpha}(\vec{r})
= \rho_{\alpha}(\vec{r}) - \rho^{SCF}_{\alpha}(\vec{r})$, about
the SCF concentration profiles ${\rho}^{SCF}_{\alpha}(\vec{r})$.
A way how to perform such a perturbative extension of the self-consistent 
field theory has recently been suggested for incompressible mixtures in
the canonical ensemble by Shi, Noolandi and Desai\cite{SHI}. 
In the following, we shall briefly sketch their approach, and generalize it
for the case of compressible mixtures and arbitrary statistical ensemble.
Capillary wave fluctuations shall be discussed subsequently.

The starting point is the partition function (\ref{zcan2}) or (\ref{zgcan2}). 
In the perturbative treatment, the free energy functional 
${\cal F}\{\rho_{\alpha}, \omega_{\alpha}\{\rho_{\alpha}\}\}$
is expanded about its self-consistent field minimum $F^{SCF}$. 
Following Shi {\em et al}, this is done most conveniently by expanding
${\cal F}_C\{\rho_{\alpha}, \omega_{\alpha}\}$ in both $\delta \rho_{\alpha}$ 
and the field deviations $\delta \omega_{\alpha} 
= \omega_{\alpha}\{\rho_{\alpha}\} - \omega_{\alpha}\{\rho^{SCF}_{\alpha}\}$, 
keeping in mind that the latter are functions of the concentration
perturbations $\delta \rho_{\alpha}$. It is useful to define the
single-chain $k$-point distribution function
\begin{equation}
\label{gg}
g^{j}_{\alpha_1 \cdots \alpha_k}(\vec{r}_1\cdots\vec{r}_k)
= \frac{1}{{\cal Q}_j}\: \frac{\delta^k {\cal Q}_j}
{\delta \omega_{\alpha_1}(\vec{r}_1) \cdots
\delta \omega_{\alpha_k}(\vec{r}_k) },
\end{equation}
which is the joint $k$-point density 
$\langle \widehat{\rho}^j_{\alpha_1}(\vec{r}_1) \cdots
\widehat{\rho}^j_{\alpha_k}(\vec{r}_k) \rangle$
for monomers of the same chain of type $j$. 
The corresponding cumulant correlation functions are given by
\begin{equation}
\label{cc}
c^{j}_{\alpha_1 \cdots \alpha_k}(\vec{r}_1\cdots\vec{r}_k)
= \frac{\delta^k \ln {\cal Q}_j}
{\delta \omega_{\alpha_1}(\vec{r}_1) \cdots
\delta \omega_{\alpha_k}(\vec{r}_k) }.
\end{equation}
For example, the lowest cumulant correlation function is simply the density,
$c^j_{\alpha}(\vec{r}) = g^j_{\alpha}(\vec{r}) = 
\langle \widehat{\rho}^j_{\alpha} \rangle $, the pair correlation function
is $c^j_{\alpha \beta}(\vec{r},\vec{r}{}') = 
\langle \widehat{\rho}^j_\alpha(\vec{r}) 
\widehat{\rho}^j_\beta(\vec{r}{}') \rangle
- \langle \widehat{\rho}^j_\alpha(\vec{r}) \rangle 
\langle \widehat{\rho}^j_\beta(\vec{r}{}') \rangle$ etc. Note that these
are single chain correlation functions of noninteracting chains $j$
subject to the external fields $\omega_{\alpha}(\vec{r})$. For
convenience, we also introduce the notation
\begin{equation}
\label{vv}
{\cal V}^{(k)}_{\alpha_1 \cdots \alpha_k}(\vec{r}_1 \cdots \vec{r}_k) = 
\frac{\delta^k {\cal V}}
{\delta \rho_{\alpha_1}(\vec{r}_1) \cdots \delta \rho_{\alpha_k}(\vec{r}_k)}
\Big|_{\{\rho_{\alpha}^{SCF}\}}
\end{equation}
and define the functions
$K_{\alpha_1 \cdots \alpha_k}(\vec{r}_1\cdots\vec{r}_k)$:
\begin{equation}
K_{\alpha_1 \cdots \alpha_k}(\vec{r}_1\cdots\vec{r}_k)
= \!\!\! \left\{ \!\!\! \begin{array}{l}
\sum_j n_j \:
c^{j}_{\alpha_1 \cdots \alpha_k}(\vec{r}_1\cdots\vec{r}_k) \quad
\mbox{
(canonical ensemble)} \\
\sum_j \exp(\beta \mu_j) \:{\cal Q}_j \:
g^{j}_{\alpha_1 \cdots \alpha_k}(\vec{r}_1,\cdots,\vec{r}_k)  \qquad
\\
 \qquad \qquad \qquad \qquad \mbox{(grand canonical ensemble). }
\end{array}
\right. 
\end{equation} 

With these definitions, and using eqns. (\ref{rcan}), (\ref{rgcan}), and
(\ref{ww}), the free energy functional (\ref{fcan}) or
(\ref{fgcan}) can be expanded as
\begin{eqnarray}
\label{ffluc1}
\lefteqn{
\beta {\cal F}\{\rho_{\alpha}, \omega_{\alpha}\} =
\beta F^{SCF}  - \sum_{\alpha} \int \!\! d \vec{r} \:
\delta \rho_{\alpha}(\vec{r}) \: \delta \omega_{\beta}(\vec{r}{}')
} \\
&& 
+ \sum_{k=2}^{\infty} \frac{1}{k!} \sum_{\alpha_1 \cdots \alpha_k} 
\int \!\! d \vec{r}_1 \cdots d \vec{r}_k \;
\; \Big\{ \;
\beta {\cal V}^{(k)}_{\alpha_1 \cdots \alpha_k}(\vec{r}_1 \cdots \vec{r}_k)\:
\delta \rho_{\alpha_1}(\vec{r}_1) \cdots \delta \rho_{\alpha_k}(\vec{r}_k)
 \nonumber \\
&& \qquad \qquad \qquad \qquad \qquad
- \; K_{\alpha_1 \cdots \alpha_k}(\vec{r}_1 \cdots \vec{r}_k) \:
\delta \omega_{\alpha_1}(\vec{r}_1) \cdots \delta \omega_{\alpha_k}(\vec{r}_k) 
\; \Big\}. \nonumber
\end{eqnarray}
The field deviations $\delta \omega_{\alpha}$ are functions of the
concentration perturbations by means of the relation (\ref{rcan}) in the 
canonical ensemble, or the relation (\ref{rgcan}) in the grand canonical 
ensemble. In terms of the correlation functions $K$ introduced above, 
these can be rewritten as
\begin{eqnarray}
\label{c-w}
\delta \rho_{\alpha}(\vec{r})& =&
- \sum_{k=2}^{\infty} \frac{1}{(k-1)!}  
\sum_{\alpha_2 \cdots \alpha_k} 
\int \!\! d \vec{r}_2 \cdots d \vec{r}_k \; \\
&&\qquad \qquad
 \: K_{\alpha \alpha_2 \cdots \alpha_k}(\vec{r} \vec{r}_2 \cdots \vec{r}_k) \:
\delta \omega_{\alpha_2}(\vec{r}_2) \cdots 
\delta \omega_{\alpha_k}(\vec{r}_k).
\nonumber
\end{eqnarray}
To leading order in $\delta \rho_{\alpha}$, one gets
\begin{equation}
\label{w-c}
\delta \omega_{\alpha}(\vec{r}) =
- \sum_{\beta} 
\int \!\! d \vec{r}{}' 
 \: K^{-1}_{\alpha \beta}(\vec{r},\vec{r}{}')
\delta \rho_{\beta}(\vec{r}{}'),
\end{equation}
where $K_{\alpha \beta}^{-1}$ is the inverse of $K_{\alpha \beta}$, 
defined through the relation
\begin{displaymath}
\sum_{\gamma} \int d \vec{r}{}''
K^{-1}_{\alpha \gamma}(\vec{r},\vec{r}{}'')
K_{\gamma \beta}(\vec{r}{}'',\vec{r}{}')
= \delta_{\alpha \beta} \delta(\vec{r}-\vec{r}{}').
\end{displaymath}
Inserting this into eqn. (\ref{ffluc1}) yields the gaussian contribution
of the fluctuations to the free energy functional
\begin{eqnarray}
\label{ffluc2}
\beta {\cal F}  &=& \beta F^{SCF}  +  \\
&& \frac{1}{2}  \sum_{\alpha \beta} \int \!\! d \vec{r} \: d \vec{r}{}'
\Big\{ 
\beta {\cal V}^{(2)}_{\alpha \beta}(\vec{r},\vec{r}{}') 
+ K^{-1}_{\alpha \beta}(\vec{r},\vec{r}{}') \Big\}
\delta \rho_{\alpha}(\vec{r}) \: \delta \rho_{\beta}(\vec{r}{}') 
+ \cdots 
\nonumber 
\end{eqnarray}
The study of these harmonic fluctuations already permits a stability
analysis of the self-consistent field solution $F^{SCF}$, and the 
calculation of scattering functions. The partition function in 
leading harmonic order contains only gaussian integrals and can be 
evaluated in a straightforward way. At first sight, inverting
the function $K_{\alpha \beta}(\vec{r},\vec{r}{}')$ seems a difficult 
task. However, a considerable simplification can often be achieved by an
appropriate basis transformation, {\em i.e.}, expressing  $K_{\alpha \beta}$, 
$\rho_{\alpha}$ etc. in terms of suitable basis functions -- 
for example, in periodically ordered structures an expansion in the
Bloch waves belonging to the periodically ordered potential
$\omega_{\alpha}(\vec{r})$ has proved useful \cite{SHI}.
Higher order corrections to eqn. (\ref{ffluc2}) can be obtained iteratively, 
yet their treatment becomes much more involved.

It is often expedient to recast the theory in a way which allows to separate 
total density fluctuations from composition fluctuations. For simplicity, we 
shall now consider a mixture which contains only two types of monomers
$A$ and $B$ of equal size. We replace the variables $(\rho_A,\rho_B)$ by
$(\rho,\phi)$ with the total density $\rho = \rho_A + \rho_B$,
and the composition variable $\phi = \rho_A - \rho_B$. The conjugate fields
are given by $\omega_{\rho} = (\omega_A + \omega_B)/2$ and 
$\omega_\phi = (\omega_A - \omega_B)/2$. Furthermore, we adopt the
definitions of Laradji {\em et al} \cite{LARADJI}
\begin{equation}
\begin{array}{lcl}
\Sigma &=& K_{AA} + K_{AB} + K_{BA} + K_{BB} \\
\Delta_1 &=& K_{AA} + K_{AB} - K_{BA} - K_{BB} \\
\Delta_2 &=& K_{AA} - K_{AB} + K_{BA} - K_{BB} \\
C &=& K_{AA} - K_{AB} - K_{BA} + K_{BB}.
\end{array} 
\end{equation}
and assume the matrix notation
$\Sigma \; \delta\omega_{\rho} = \int \!\! d \vec{r}{}' 
\Sigma (\vec{r},\vec{r}{}') \delta \omega_{\rho}(\vec{r}{}') $.
Equation (\ref{c-w}) can then be rewritten to leading gaussian order
\begin{equation}
{\delta \rho \choose \delta \phi}  = 
\left( \begin{array}{ll} 
\Sigma  & \Delta_2  
\\ \Delta_1 & C  \end{array} \right) \:
{\delta \omega_{\rho}\choose \delta \omega_{\phi}} + \cdots
\end{equation}
Inverting this expression yields
\begin{equation}
{\delta \omega_{\rho}\choose \delta \omega_{\phi}}  
= \left( \begin{array}{ll} 
\tilde{\Sigma}^{-1}  & 
\tilde{\Delta}_1^{-1}  \\
\tilde{\Delta}_2^{-1} & 
\tilde{C}^{-1}  \end{array} \right) 
{\delta \rho\choose \delta \phi}  
 + \cdots 
\qquad \qquad \qquad \qquad \qquad \qquad
\end{equation}
\begin{equation}
\qquad \qquad \qquad \qquad \mbox{with} \qquad
\begin{array}{lclcl}
\tilde{\Sigma} &=& \Sigma &-& \Delta_2 C^{-1} \Delta_1 \\
\tilde{\Delta}_1 &=& \Delta_1 &-& C \; \Delta_2^{-1} \Sigma \\
\tilde{\Delta}_2 &=& \Delta_2 &-& \Sigma \; \Delta_1^{-1} C \\
\tilde{C} &=& C &-& \Delta_1 \Sigma^{-1} \Delta_2  \quad .
\end{array} 
\end{equation}
The free energy expansion (\ref{ffluc2}) thus takes the form
\begin{eqnarray}
\label{ffluc3}
\beta {\cal F} &=& \beta F^{SCF}  + \\
&& \frac{1}{2}  \:
(\delta \rho \; \; \delta \phi) 
\left( \begin{array}{l @{\qquad} l} 
\tilde{\Sigma}^{-1} \!\!  + \beta {\cal V}_{\rho \rho}^{(2)} & 
\tilde{\Delta}_1^{-1} \!\!+ \beta {\cal V}_{\rho \phi}^{(2)}  \\
\tilde{\Delta}_2^{-1} \!\! + \beta {\cal V}_{\phi \rho}^{(2)} &  
\tilde{C}^{-1} \!\! + \beta {\cal V}_{\phi \phi}^{(2)} 
\end{array} \right)
{\delta \rho \choose \delta \phi} 
+ \cdots,  \nonumber
\end{eqnarray}
where ${\cal V}^{(2)}_{\phi \phi} (\vec{r},\vec{r}{}')$ is defined as
$\delta^2 {\cal V}/\delta \phi(\vec{r})\: \delta \phi(\vec{r}{}')$ etc.,
in analogy to eqn. (\ref{vv}). Usually in the 
literature\cite{SHI,LARADJI,MARK3}, 
incompressible melts have been considered \mbox{($\delta \rho = 0$)}
with an interaction potential of the form 
$\beta {\cal V}_{\mbox{\tiny inter}}=
\int\! d\vec{r}\:\chi \:(\rho_A \rho_B)/\rho$.
In this case, the free energy can be written as
\begin{equation}
\beta {\cal F} = \beta F^{SCF} +
\frac{1}{2}  \:
\delta \rho \: [C^{RPA}]^{-1} \delta \rho
\quad \mbox{with} \quad
C^{RPA} = [\tilde{C}^{-1} - \frac{\chi}{2} \mbox{\bf 1}]^{-1}.
\end{equation}
Equation (\ref{ffluc3}) generalizes this result for
arbitrary potentials ${\cal V}$ and compressible blends.

In well segregated macro- or microseparated blends, the main composition 
fluctuations are due to capillary wave fluctuations of interface positions.
These are less well captured by perturbative treatments, since the local
concentration deviations $\delta \rho_{\alpha}(\vec{r})$ are not small. As
long as the interfaces are still localized, {\em i.e.}, in periodic
structures, the method can still be applied to some extent\cite{SHI}.
However, the expansion will fail if the fluctuations are strong enough to 
destroy the long range order. Note that single, isolated
interfaces between macroscopic phases usually delocalize 
\cite{capillary,SHULL2,SEMENOV1,TOBIAS,JONES}, 
{\em i.e.}, the extent of the fluctuations depends on the size of the system. 
In order to study these long wavelength fluctuations, other approaches like 
effective interface descriptions with input parameters taken from 
self-consistent field calculations are more appropriate. 

In order to illustrate the problem, we shall discuss the simplest
example, a single interface separating two macroscopic phases in a blend of
incompatible, but otherwise symmetric $A$ and $B$ homopolymers. 
Let us assume  that is only slightly distorted from its lowest energy flat 
state, such that the local deviations of the interfacial position can be 
described by a single valued function $h(x,y)$, and the gradients 
$|\nabla h|$ are small. 
The increase in interfacial area caused by these fluctuations
cost the free energy
\begin{equation}
\label{cw}
{\cal H}_{CW} = \frac{\sigma}{2} \int \! \! dx \: dy \: |\nabla h|^2,
\end{equation}
where $\sigma$ is the interfacial tension, and energy losses
due to the distortion of profiles and the like have been neglected. 
The functional (\ref{cw}) is commonly referred to as the capillary wave 
Hamiltonian. It can be diagonalized by means of a Fourier transformation 
in $x$ and $y$, and since it is quadratic, the spectrum and the distribution 
functions can be determined analytically.
The thermal average of the Fourier components is given by
\mbox{$\langle |h(\vec{q})|^2 \rangle = k_B T /(\sigma q^2)$}
and one obtains a Gaussian height distribution function\cite{capillary}
\begin{equation}
\langle \delta( z - h(x,y)) \rangle = P_{s^2}(z) = 
\frac{1}{2 \pi s^2} \exp(-\frac{z{}^2}{2 s^2})
\end{equation}
\begin{equation}
\mbox{with} \qquad 
s^2 = \frac{1}{4 \pi^2} \int d\vec{q} \; \langle | h(\vec{q}) |^2 \rangle =
\frac{1}{2 \pi \sigma} \ln(\frac{q_{\mbox{\tiny max}}}{q_{\mbox{\tiny min}}}).
\end{equation}
Here one has to introduce an upper cutoff $q_{\mbox{\tiny max}}$ 
and a lower cutoff $q_{\mbox{\tiny min}}$, since the integral 
$\int dq/q$ diverges both at $q \to \infty$ and $q \to 0$. The lower cutoff 
is obviously given by the system size, $q_{\mbox{\tiny min}}= 2 \pi/L$. 
An important consequence is that the width $s$ of the distribution function
$P(h)$ grows logarithmically with the system size $L$, {\em i.e.}, the
interface is marginally rough. 

The value of the upper cutoff is less obvious. Clearly, the 
capillary wave Hamiltonian (\ref{cw}) cannot be expected to provide a good 
description of polymer interfaces on all length scales. Ideally, one would 
hope that one can find a microscopic length $1/q_{\mbox{\tiny max}}$ beyond 
which (\ref{cw}) is valid, and that the system can be studied independently
by other means on smaller length scales. This implies that the coupling of 
the long wavelength capillary wave fluctuations with the local structure on 
short length scales can be neglected. 
In that case, an approximation makes sense which describes the interfacial
structure by local ``intrinsic'' profiles, centered at the local interface
position, \mbox{$\rho_{\alpha}(x,y,z) = \rho_{\alpha}^{(int)}(z - h(x,y))$}
(convolution approximation\cite{JASNOW}). 
The intrinsic profile characterizes the system on the length scale of 
$1/q_{\mbox{\tiny max}}$. When looking at the interface on a larger
length scale $1/q_0$, one obtains apparent profiles 
$\rho_{\alpha}^{(app)}(z)$, which are broadened by the capillary wave
modes with wavevectors between $q_0$ and $q_{\mbox{\tiny max}}$,
\begin{equation}
\label{conv}
\rho_{\alpha}^{(app)}(z) = \int_{-\infty}^{\infty} \!\!\!
dh \: \rho_{\alpha}^{(int)}(z-h) \: P_{s'^2}(h)
\quad \mbox{with} \quad
s'^2 = \frac{1}{2 \pi \sigma} \ln(q_{\mbox{\tiny max}}/q_0).
\end{equation}
Note that since \mbox{$P_{s^2}(z) = \int_{-\infty}^{\infty} \!\!\!
dh \: P_{s'^2}(z-h) \: P_{(s^2-s'^2)}(h)$}, the intrinsic profile 
$\rho_{\alpha}^{(int)}(z)$ can be replaced by $\rho_{\alpha}^{(app)}(z)$, 
and the upper cutoff $q_{\mbox{\tiny max}}$ by $q_0$, without affecting
anything on length scales beyond $1/q_0$. 
Thus the choice of the upper cutoff is largely arbitrary.

\begin{figure}

\fig{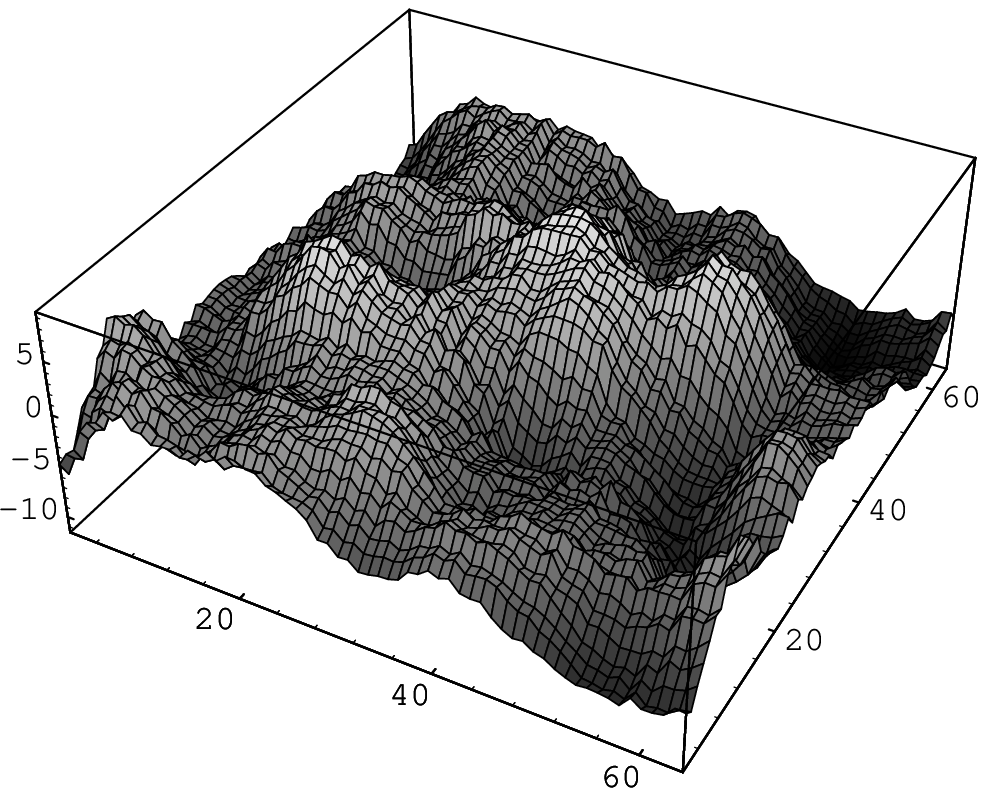}{100}{80}

\caption{\label{snapshot}}
Typical snapshot picture of the local interface position $h(x,y)$ in
Monte Carlo simulations of a symmetric homopolymer interface at 
$\chi=0.16, N=32$. The coarse-graining length $B=8$ is roughly the chain's 
gyration radius. System dimensions are $D=64$ and $L=64$. 

from Ref. \cite{ANDREAS2}

\end{figure}

\begin{figure}

\fig{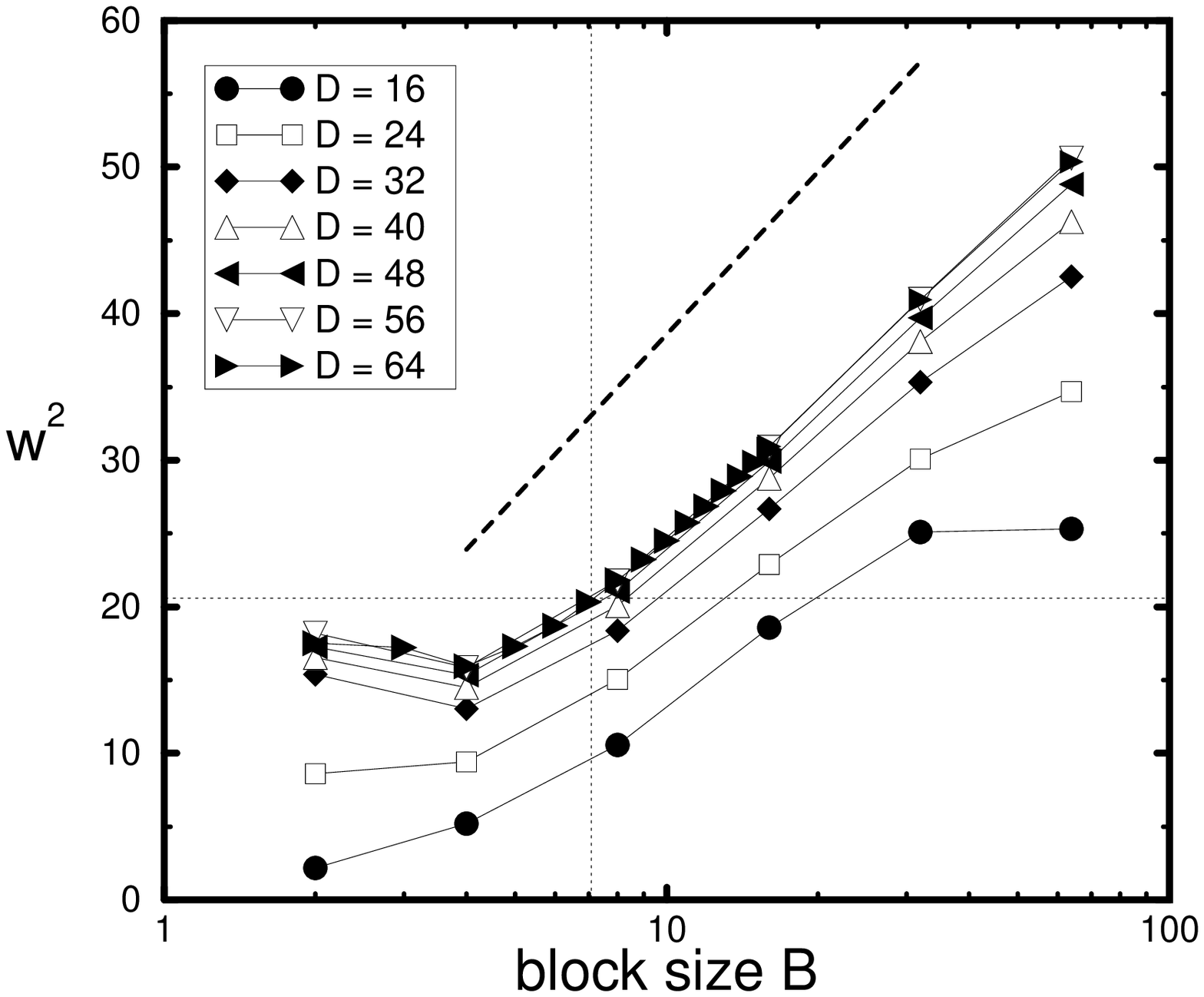}{100}{80}

\caption{\label{w2_b}}
Squared interfacial width $w^2$ as a function of block size $B$ in 
Monte Carlo simulations of a homopolymer interface confined in films
of various thicknesses $D$ with lateral dimension $L=128$,
at $\chi=0.16$ and $N=32$. The horizontal dotted line marks the self 
consistent field prediction for a free interface $w^2 = 20.5$, and the 
vertical dotted line the value of the gyration radius of a chain $R_g = 7.05$.
The dashed line indicates the theoretically predicted slope of 
$1/(4 \sigma) \; \ln(B)$, with $\sigma=0.0156$ taken from the self 
consistent field theory.

From Ref. \cite{ANDREAS2}
\end{figure}

Werner {\em et al}\cite{ANDREAS2} have tested this concept by extensive Monte 
Carlo simulations of interfaces between immiscible phases in symmetric binary 
polymer blends, within the bond fluctuation model (see Section 3.1).
The simulations were done in a $L\times L\times D$ geometry 
with periodic boundary conditions in the $L$ directions, and hard walls in 
the $D$ direction, which favor one the $A$ component and the other
the $B$ component. The wall interaction parameters were chosen beyond
the wetting transition, hence large enough to enforce a delocalized
$AB$ interface which is on average located in the middle of the film. 
In thin films, the capillary wave fluctuations are limited by the film 
thickness $D$ rather than by the system size $L$. However, if the film
thickness $D$ is chosen large enough compared to $L$, the interface is
essentially free. In order to study the interfacial fluctuations,
the system was split into columns of block size $B \times B$ and height
$D$, and the Gibbs dividing surface $h(x,y)$ was determined in each
column. A typical snapshot of the resulting local interface position
$h(x,y)$ is shown in Figure \ref{snapshot} for the block size $B=8$ lattice
constants, which is roughly the radius of gyration ($R_g \approx 7$).
The monomer profiles were then taken relative to this position, and after
averaging over all columns, the interfacial width $w$ was determined
by fitting the order parameter profile 
$m(z) = (\rho_A(z) - \rho_B(z))/\rho(z)$ to a $\tanh$ profile,
$m(z) = m_b \tanh(z/w)$. Using eqn. (\ref{conv}), one can show\cite{ANDREAS2} 
that the apparent width obtained with this procedure is broadened according to
\begin{equation}
\label{wbroad}
w^2 = w_0^2 + \frac{1}{4 \sigma} \ln (\frac{B}{B_0})
\end{equation}
due to capillary waves, where $w_0$ is the intrinsic width and
$B_0=2 \pi /q_{\mbox{\tiny max}}$ is the coarse-graining length associated  
with the upper cutoff $q_{\mbox{\tiny max}}$. Figure \ref{w2_b} shows the 
simulation results for the squared interfacial width $w^2$ as a function of 
block size $B$ in films of lateral size $L=128$, for various film thicknesses 
$D$. They become independent of $D$ for thicknesses larger than $D > 48$, 
hence the interface can then be considered to be free.
For large block sizes $B$, $w^2$ grows logarithmically with $B$,
with a slope which is in rough agreement with the theoretical prediction
$1/(4 \sigma)$ (dashed line). For very small block sizes $B \le 4$,
$w^2$ becomes flat. The self-consistent field value $w_{SCF}{}^2 = 20.9$
is reached at the block size $B_0 \approx 7$.
However, looking just at Figure \ref{w2_b},
nothing in the shape of the curves indicates that there should be
anything special about the block size 7 or about $w^2=20.5$.
The regime of logarithmic growth already
starts at much smaller block sizes. On the other hand, $w_{SCF}$ can be
made intrinsic width by decree. With the upper cutoff $q_{\mbox{\tiny max}}$ 
defined by the corresponding block size $B_0$, the capillary wave 
Hamiltonian then provides a reasonable description of the simulation data.
The question remains which is the correct choice of the cutoff $B_0$.
In our example, $B_0$ happens to be exactly the gyration radius of the 
chain, but also twice the statistical segment length of a chain ($b=3.05$), or
one and a half times the interfacial width. Each of these is a valid
candidate for the cutoff parameter. Semenov\cite{SEMENOV1}, for example,
favors the interfacial width. Here again simulations can help to
clarify the issue. Recently, Werner {\em et al} have repeated the
analysis leading to Fig. \ref{w2_b} with systematically varied chain 
length $N$ and/or the monomer interactions, {\em i.e.}, the $\chi$ parameter. 
At fixed $\chi N=5.1$, they find that the cutoff $B_0$, defined by 
$w(B_0)=w_{SCF}$, scales like 
$B_0 \propto \sqrt{N} \propto \sqrt{1/\chi} \propto w_{SSL} \propto R_g$.
At fixed $\chi = 0.16$, $B_0$ first increases strongly with $N$, but
levels off faster than $R_g \propto \sqrt{N}$ at the largest chain
lengths ($N=256$) \cite{ANDREASD,ANDREAS3}.

In order to gain a better understanding of this problem, let us go back to
the case of infinitely long chains and recall that self-consistent field
theory is supposed to be exact in this limit. More specifically, it gives
the correct free energy of the system. On the other hand, the interface has
a given, finite surface tension ({\em e.g.}, $\sigma_{SSL}$ as obtained in
\cite{HELFAND1}), hence the interface position will fluctuate.
Concentration fluctuations may exist in the 
long chain limit, they just do not affect the thermodynamics of the system. 
However, they influence the local structure, such as local concentration 
profiles or orientational properties at interfaces. Self consistent field 
calculations can thus not be expected to give equally good results on all 
length scales. When used to study local structure properties in polymer 
mixtures, one has to ask for the length scale for which the theories describe 
the system best. 
The simulations of Werner {\em et al} suggest that it probably approaches a
constant $(B_0 \to (3-4) \; w_{SSL})$ in the strong segregation limit
$(N \to \infty)$, but is subject to strong chain end corrections.
It is worth noting that the cutoff $B_0$ is always much larger than the
block size at which $w^2(B)$ starts to grow logarithmically
as predicted by (\ref{wbroad}). Up to the length scale of $B_0$, the
structures obtained with self-consistent field theories thus average
not only over the bulk composition fluctuations, but also over the
capillary wave fluctuations of the interface position.

\section{Applications}

Self consistent field theories are finding widespread use in numerous
contexts of polymer or macromolecular physics. For example, they
have been employed to calculate complicated phase diagrams of copolymer
blends or mixtures of copolymers and homopolymers\cite{MARKR,PHILIP2}, and 
to study details of density profiles at the internal interfaces in such
materials\cite{SHULL1,SHULL2,NOOLANDI2,FISCHEL,BALASZ1}. 
They are frequently applied in surface physics,
{\em e.g.}, in connection with surface segregation\cite{GENZER,HARIHARAN,FS3} 
and wetting phenomena\cite{MARCUS6}, or for the theoretical study of polymers 
or copolymers which are adsorbed on surfaces or grafted to 
surfaces\cite{FLEERB,FLEERR,BALASZ2}.
We shall not attempt to give an account of all these activities here.
Rather, we refer the reader to the various recent reviews on the
different topics, {\em e.g.}, Refs. \cite{FLEERB,SZLEIFERR,MARKR,SANCHEZ},
and illustrate the use of the method with two examples which are taken from 
the present author's own research: A comparison between self-consistent
field calculations and Monte Carlo simulations of polymer conformations
at interfaces, and a self-consistent field study of the phase behavior 
of short amphiphilic molecules at surfaces.

\subsection{Conformations of polymers at interfaces}

We consider an interface between coexisting phases in a symmetric mixture of 
homopolymers $A$ and $B$, in the strong segregation regime ($\chi N = 17$),
{\em i.e.}, at temperatures well below the demixing temperature.
It is studied by self-consistent field theory within the Gaussian chain 
model and the wormlike chain model.

The self-consistent field calculations are compared with simulations of the 
bond fluctuation model. The latter represents polymers by chains of
spatially extended effective monomers, which occupy each a cube of
eight neigboring sites on a cubic lattice, and which are connected by
bonds of length $\le \sqrt{10}$ lattice spacings. The two types of
monomers $A$ and $B$ interact pairwise with interactions 
$\epsilon_{AA} = \epsilon_{BB}=-\epsilon_{AB} = -k_B T \epsilon$, if they 
are less than $\sqrt{6}$ lattice units apart. The equation of state and the 
local pair correlations in the bulk are well known. Moreover, the model was 
shown to behave like a dense melt at volume fraction 0.5 or density 
$\rho=1/16$, {\em i.e.}, the chains have almost ideal Gaussian statistics,
with known statistical segment length $b \approx 3$ lattice units.
The simulations were performed at chain length $N=32$ and interaction
strength $\epsilon=0.1$, which corresponds to $\chi = 0.53$.

\begin{figure}

\fig{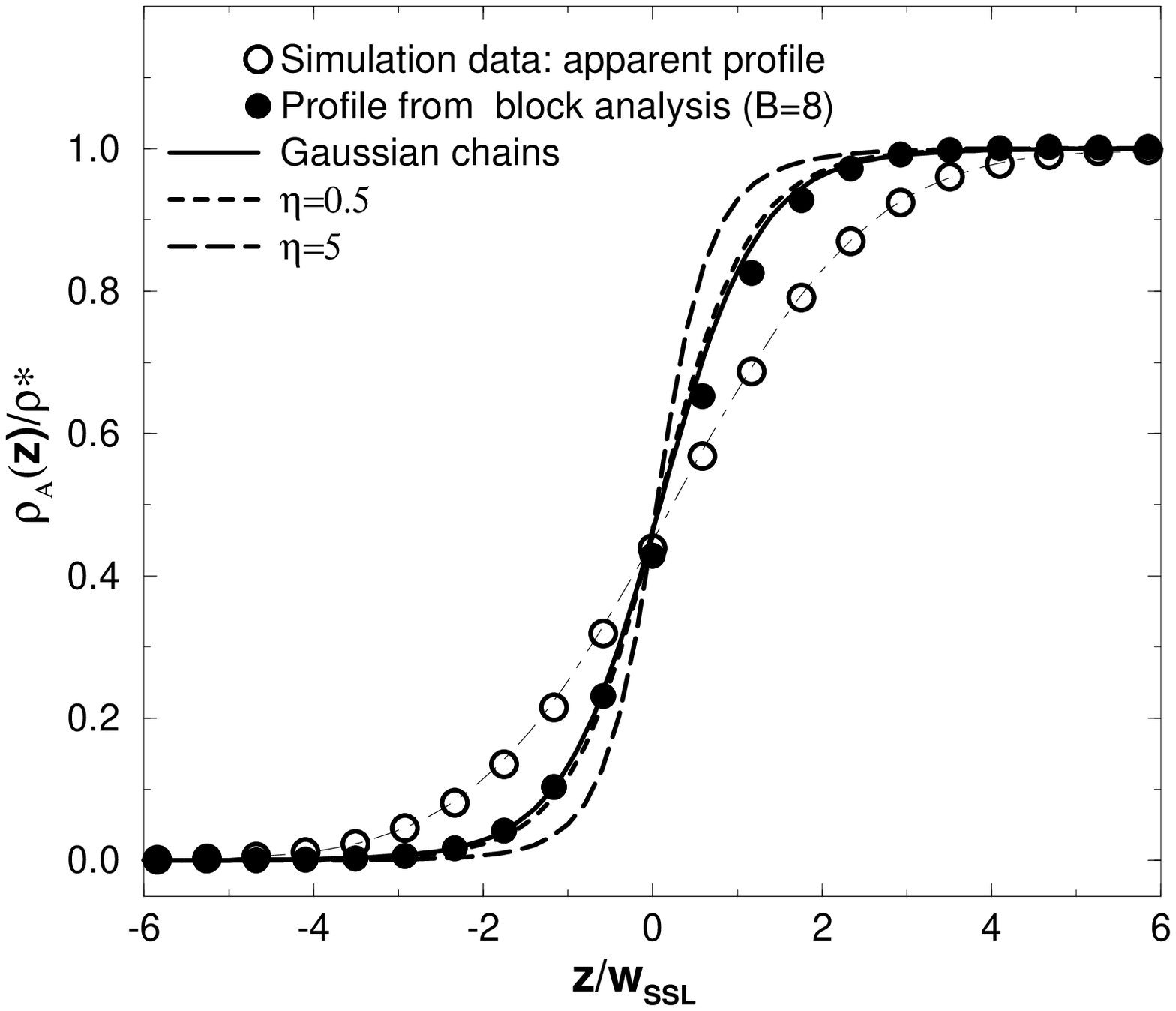}{100}{80}

\caption{\label{rhoa}}
Concentration profile $\rho_A/\rho^*$ vs. $z/w_{SSL}$ (with
$w_{SSL}=b/\sqrt{6 \chi }$) in self-consistent field theory for 
Gaussian chains (solid line) and wormlike chains of
different chain stiffness $\eta$ (dashed lines), at
$\chi = 0.53$ and chain length $N=32$. 
The results are compared to simulation data without (open circles) 
and with (closed circles) block splitting. The thin dashed dotted
line shows the profile for $\eta=0.5$ with capillary wave
correction, assuming $s^2 = 2.5$ (see eqn. \ref{conv})\cite{FN2}.

From Ref. \cite{FS1} and \cite{ANDREASD}.
\end{figure}

Figure \ref{rhoa} shows the results for the interfacial profile
(\cite{FS1} and \cite{ANDREASD}). All lengths are given in units of
$w_{SSL}=b/\sqrt{6 \chi}$, which is the interfacial width in an
incompressible mixture of infinitely long polymers\cite{HELFAND1}, 
and a ``natural'' unit of length within the self-consistent field theory. 
In these units, the self-consistent field theory predicts that 
the interfacial thickness decreases with increasing chain stiffness $\eta$. 
However, the statistical segment length $b = a \sqrt{2 \eta}$ 
increases in turn, such that the net effect is positive: 
In absolute units ({\em e.g.}, units of the monomer size $a$), 
the interfacial width increases with the chain stiffness $\eta$ in
this regime of relatively small chain stiffness $\eta$. (Note that
Morse and Fredrickson predict the opposite effect, a decrease of
the interfacial width, in the limit of large $\eta$.

We turn to the comparison with the Monte Carlo data. The normalization
in units of $w_{SSL}$ still makes sense, since $b$ and $w_{SSL}$ are
known, whereas the parameter $\eta$ is not. The bare simulation
profile is broader by a factor of almost two than the self-consistent
field prediction (open circles). However, this can basically be traced
back to the effect of capillary waves, as discussed in section 2.4.
The simulation profiles obtained after splitting the system into blocks of 
size $B=8\approx R_g$ as in Ref. \cite{ANDREAS2} (where the much
weaker segregated case $\chi=0.16$ was studied)
are in good agreement with the self-consistent field prediction 
(closed circles)\cite{FS1}. Alternatively, one can also correct
the self-consistent field profile for capillary wave broadening 
by means of eqn. (\ref{conv})\cite{FN2}.

Next, the conformations of polymers in the vicinity of such an interface are 
analyzed. We will examine the conformations of the constituting homopolymers, 
and those of single symmetric $A:B$ diblock copolymers (of same length), 
which adsorb to the interface. The results can again be compared with Monte 
Carlo simulations of very diluted copolymers at a homopolymer 
interface\cite{ANDREAS1}. We will limit ourselves to the discussion of 
the orientational properties of the molecules here. 

It is instructive to study separately the orientations of single bonds,
of chain segments, and of whole chains. 
In self-consistent field theory, bond orientations are conveniently
calculated within the wormlike chain model. In the following, we
will set $\eta=1/2$. At that stiffness, the statistical segment length
$b$ is identical to the ``monomer length'' $a$, hence adjacent monomers
are essentially uncorrelated. This assumption seems reasonable for
the bond fluctuation model, as long as no bond potentials have been
introduced. (An improved guess for $\eta$ would probably be slightly 
larger than $1/2$, since chains cannot fold back onto themselves). 
Note that chains with such a small stiffness behave almost like
Gaussian chains (cf., {\em e.g.}, Fig. \ref{rhoa}).

\begin{figure}

\fig{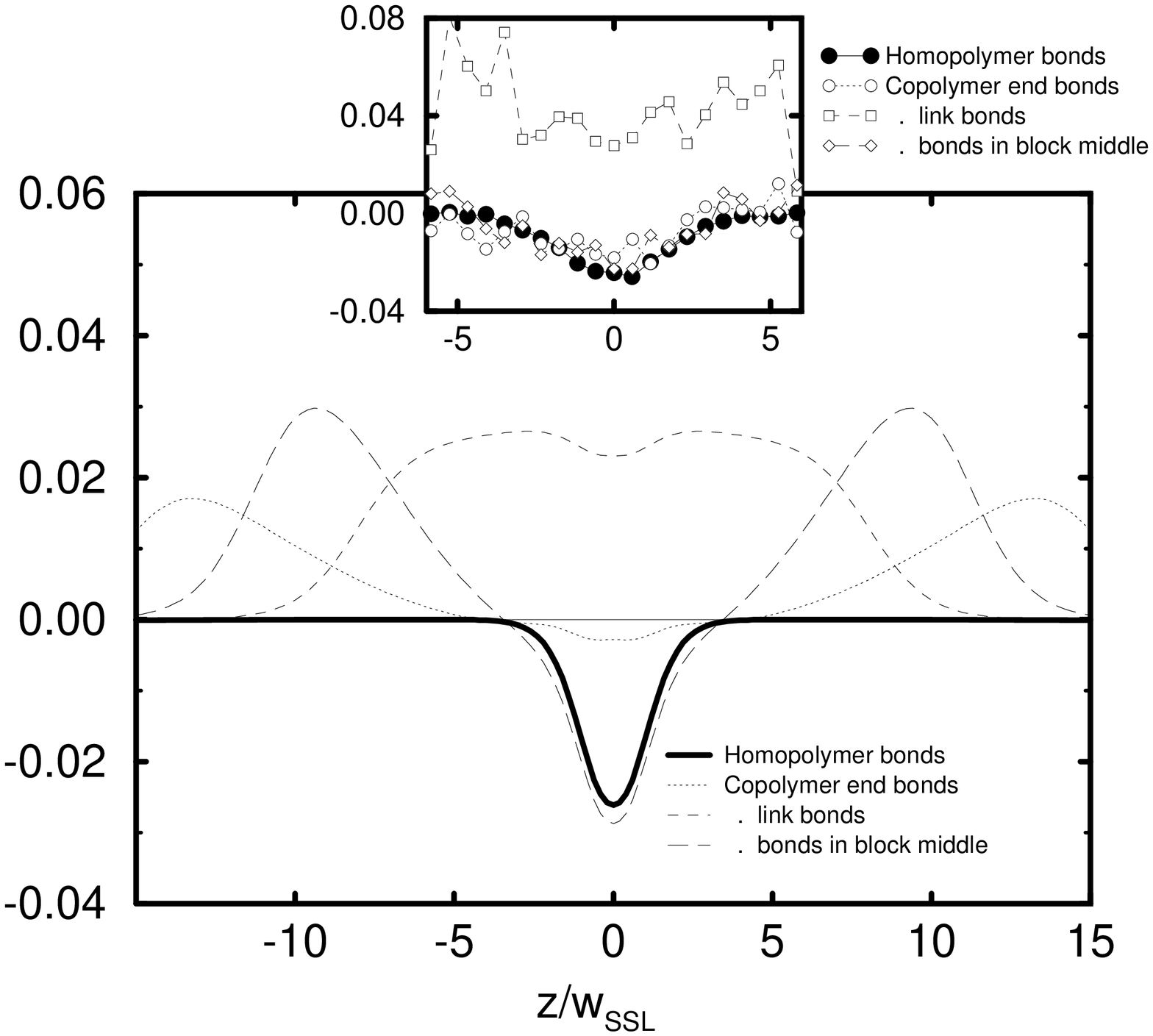}{100}{80}

\caption{\label{qb}}
Orientational order parameter $q$ vs. $z/w_{SSL}$ at a homopolymer interface
($\chi=0.53$, $N=32$) for homopolymer bonds (thick solid line) and for bonds 
in an adsorbed diblock copolymer at different positions within the chain:
end bonds (dotted), bonds linking the two blocks (dashed), bonds in
the middle of a block (long dashed). The inset shows corresponding 
Monte Carlo data.

From Ref. \cite{FS1} and \cite{ANDREAS1}.
\end{figure}

In order to study profiles of bond orientations $\vec{b}$, we
define the bond orientation parameter
\begin{equation}
q(\vec{r}) = \frac{\langle b_z{}^2 \rangle - 
\frac{1}{2} (\langle b_x{}^2 \rangle + \langle b_y{}^2 \rangle)}
{\langle \vec{b}^2 \rangle},
\end{equation}
which is negative for orientation parallel to the interface, and positive for 
perpendicular orientation. The interfacial profiles of $q$ are shown
in Figure \ref{qb}. Orientation effects on the bond level are found
to be overall very weak. Homopolymer bonds tend to orient parallel to the
interface in the interfacial region. The same holds for most parts of
the copolymer; Only very few bonds in the central region connecting
the $A$ and $B$ block turn perpendicular to the interface. For both
copolymers (dotted line) and homopolymers (not shown), the orientation is 
weakest at the chain ends. These results are basically in agreement with
the Monte Carlo data (inset)\cite{ANDREAS1}. 

Further away from the interface, the orientation of copolymer monomers
is driven by a different effect. The chains are pulled towards the interface 
by one copolymer end, and the bonds align perpendicular as a result. 
At distances of several radii of gyration, the copolymers loose contact to 
the interface and the orientation parameter $q$ drops back to zero. 
The profile of $q$ for copolymers thus reflects two different length scales 
-- a tendency of parallel alignment in a region of the extent of the 
interfacial width, which gives rise to the central dip in the profiles, 
and a force towards perpendicular alignment over the length scale of the 
gyration radius, which is the range of the interaction between the copolymer 
and the interface.
Note however that most copolymer monomers are located close to the
center of the the interface, and the net bond orientation is parallel. 

\begin{figure}

\fig{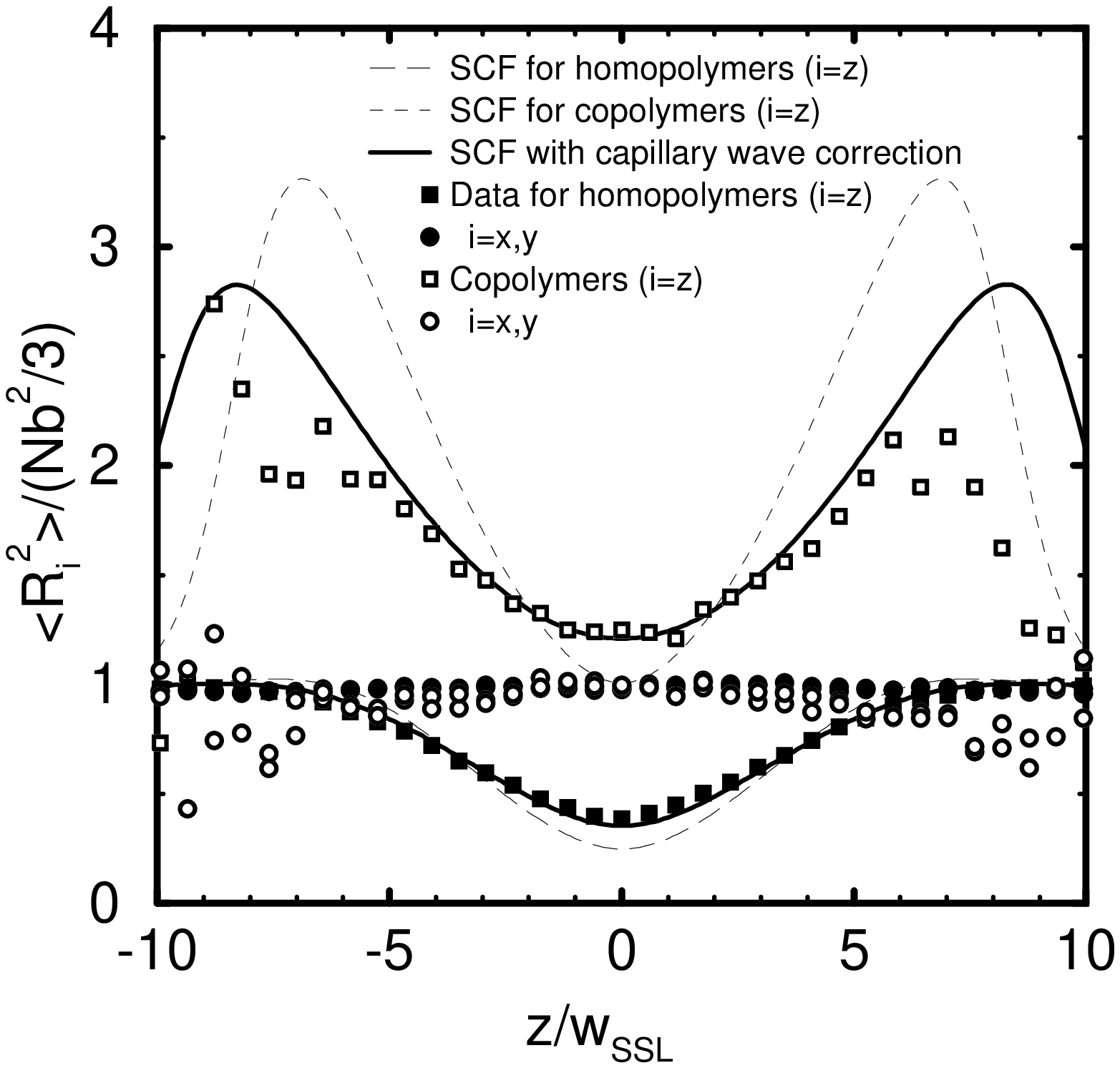}{100}{80}

\caption{\label{rgz2t}}
Mean square end-to-end vector component $\langle R_i{}^2 \rangle$ for
$i=x,y,z$ in units of the average bulk value $b^2 N/3$, vs. the
distance $z$ of the center of the end-to-end vector from the
interface in units of $w_{SSL}$, for homopolymers (long dashed) and
single copolymer chains (dashed). Parameters are $\chi=0.53, N=32$.
Points are the corresponding Monte Carlo data. The thick solid line 
gives the self-consistent field prediction with capillary wave correction
according to eqn. \ref{conv}, with $s^2=2.5$ taken from Fig. \ref{rhoa}.

From Ref. \cite{ANDREAS1} 
\end{figure}

\begin{figure}

\fig{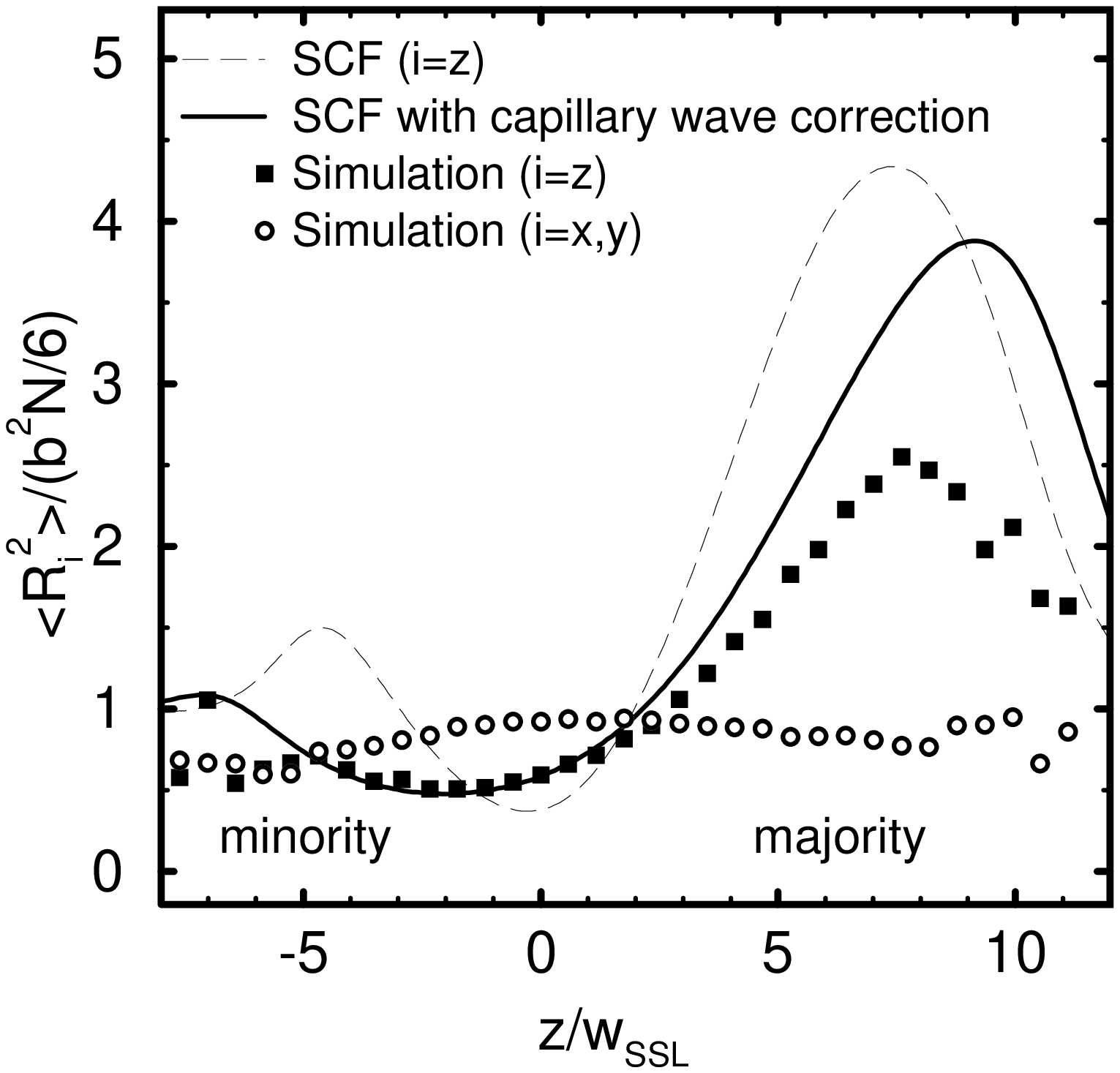}{100}{80}

\caption{\label{rgz2b}}
Mean square end-to-end vector component $\langle R_i{}^2 \rangle$ ($i=x,y,z$) 
of the copolymer blocks in their minority phase ($A$ block in $B$ phase and 
vice versa) and in their majority phase ($A$ in $A$, $B$ in $B$),
in units of the average bulk value $b^2 N/6$, plotted vs. the
distance $z$ of the center of the end-to-end vector from the
interface in units of $w_{SSL}$, compared to Monte Carlo data.
Parameters are $\chi=0.53, N=32$. Thick solid line shows the
self-consistent field prediction corrected for capillary waves
according to eqn. \ref{conv}, with $s^2=2.5$ taken from Fig. \ref{rhoa}.

From Ref. \cite{ANDREAS1}.
\end{figure}

Even though interfaces orient single bonds only very weakly, their effect
on whole chains is much stronger. The orientation of whole chains
involves two different factors: The orientation of the gyration tensor
at constant total gyration radius or end-to-end radius,
and  stretching or compression of the chain in one direction.
The first factor can not be assessed in a self-consistent field
calculation with Gaussian or nearly Gaussian chains. However, the
simulation data shown in Fig. \ref{rgz2t}\cite{ANDREAS1} indicate that
the second effect dominates close to an interface: The mean-squared
components parallel to the interface $xy$ of the end-to-end vectors hardly
vary throughout the system, for both homopolymers and copolymers,
whereas they strongly depend on the distance from the interface for
the $z$ component. Homopolymers are found to be squeezed towards
the interface, which leads to an effective parallel orientation.
Copolymers show the inverse behavior, they stretch in the direction
perpendicular to the interface. The effect is strongest for copolymers
centered between one and two radii of gyration away from the interface,
which are pulled towards the interface by their one end, and much
weaker for copolymers centered right at the interface, which do not feel 
strong orienting forces. The latter can be pictured as consisting of
two weakly coupled, almost unperturbed homopolymer blocks $A$ and $B$. 
Indeed, the end-to-end vectors of single blocks centered at the interface
are oriented parallel to the interface, as shown in Fig. \ref{rgz2b}.

To conclude, these examples demonstrate that the self-consistent field theory
allows to study the conformational properties of polymers at
interfaces in great detail, and that a wealth of information can be
obtained from such calculations. The predictions of the theory were
found to be in overall good agreement with Monte Carlo simulations.

\subsection{Amphiphiles at surfaces}

Our second example deals with a somewhat more exotic application of a
self-consistent field theory, the study of a coarse-grained model for 
Langmuir monolayers\cite{FS2,FS4}. 
These are monolayers of amphiphilic molecules adsorbed 
on a water surface. If the nonpolar chains of the amphiphiles are 
sufficiently long, one observes experimentally two distinct coexistence 
regions between two dimensional fluid phases on increasing the area per 
molecule\cite{langmuir}:
A transition from a highly diluted, ``gas'' like phase (G) 
into a more condensed, ``liquid expanded'' (LE) phase, and a second region at 
higher surface coverage, where ``liquid condensed'' domains are present in a 
``liquid expanded'' environment. The coexisting high density phases are true 
fluid phases, as positional correlations within them decay
exponentially within a few nanometers. 
In contrast, the directions of the bonds connecting nearest neighbor head 
groups appear to be correlated over tens of micrometers\cite{HELM}, 
which suggests that those phases are probably hexatic.
They may be untilted (LS) or tilted (denoted L${}_2$ here), with different
directions of tilt.
The transition between the liquid expanded and the liquid condensed phase
is the monolayer equivalent to the ``main transition'' in bilayers, 
where the bilayer thickness jumps discontinuously as a function 
of temperature \cite{main}. The latter is probably relevant in biological 
systems, because it is found in lipid membranes at temperatures often close 
to the body temperature ({\em e.g.}, $41.5^0C$ in DPPC).

The problem we wish to address is the following: What is the origin
of the first order transition between these two fluid phases?
The onset of bond orientational order cannot account for the 
discontinuity, since the transition between a liquid and a hexatic liquid 
is of Kosterlitz-Thouless type\cite{HALPERIN} and thus continuous.
There is numerous experimental evidence that the flexibility of 
the polar tails plays a crucial role. In particular, the liquid expanded 
phase disappears if the chains are made stiff by replacing the hydrogene
atoms with fluorine atoms\cite{RICE}. The system is thus a good candidate 
for a self-consistent field treatment, which takes due account of the
conformational degrees of freedom of the chains.

\begin{figure}

\fig{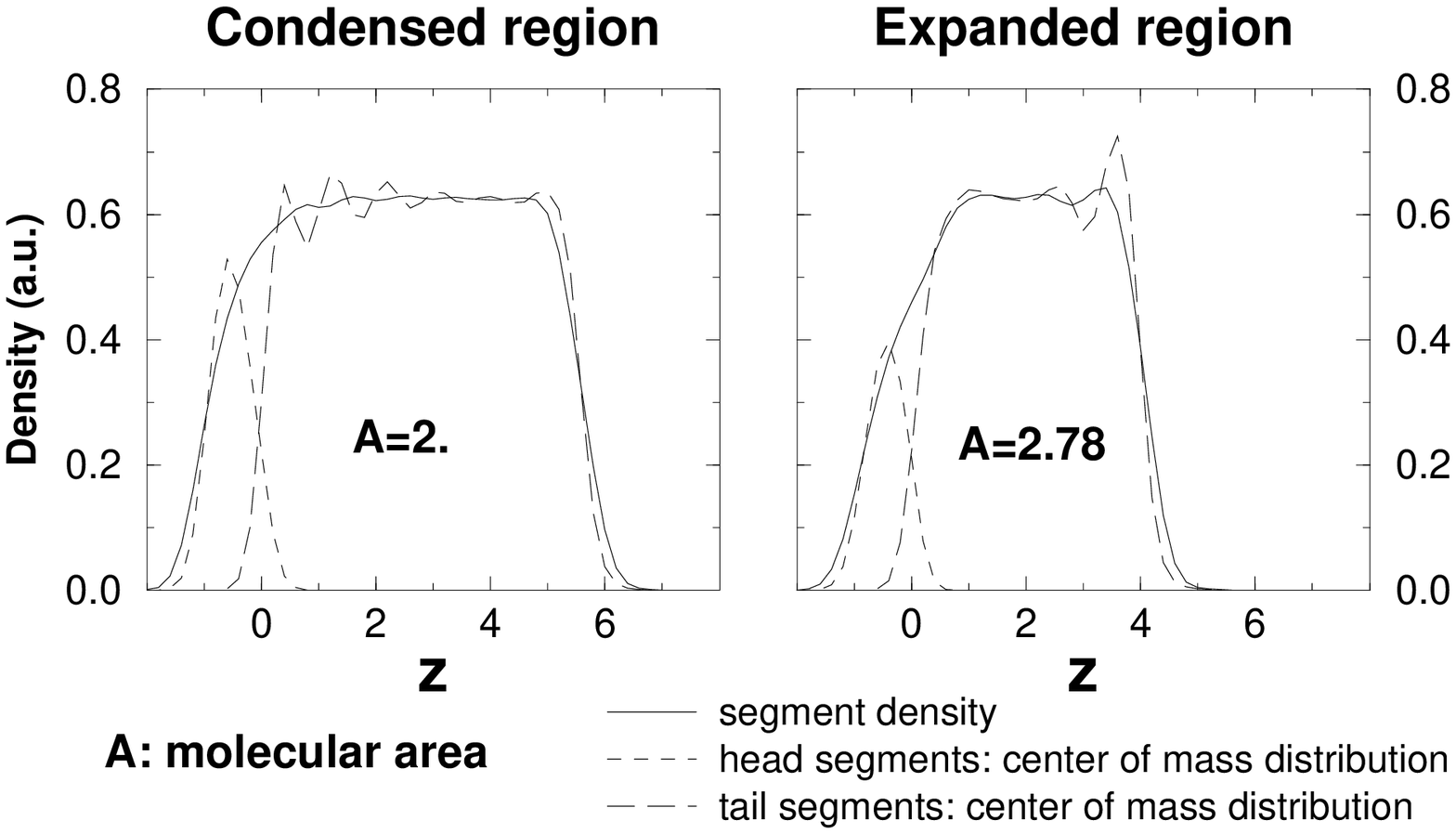}{100}{80}

\caption{\label{prof-13.7}}
Density profiles $\rho$ in units of $1/(A_0 l_0)$ vs. $z$ in units of $l_0$ 
for different molecular areas $A/A_0$. Long and short dashed lines
show the center of mass densities of tail and head segments, 
respectively. Solid line shows the coarse-grained density, which
accounts for the finite extension of the segments. Parameters are
$u=2$ and $v=13.7$.

From Ref. \cite{FS2}.
\end{figure}

The amphiphiles were modeled as chains containing one head segment,
which is confined to a planar surface by a harmonic potential, and
seven tail segments of diameter $A_0$ and length $l_0$.
The conformational weight is given by an expression of the type
(\ref{pjoint}) with an additional bending stiffness contribution
$\exp[u \widehat{U}(\theta)]$, which favors parallel
alignment of adjacent segments ($\theta = 0$). The adjustable
parameter $u$ determines the stiffness of the chains, and the actual form 
of $\widehat{U}$\cite{FS2} is not of interest here. Chain
segments interact {\em via} repulsive hard core and long range attractive
forces. The self-consistent field treatment of the interactions essentially 
follows the lines of Section 2.3, except that in those short chains, 
one has to account explicitly for the extended size the monomers by 
some appropriate coarse-graining over the center-of-mass densities 
of segments.
Furthermore, the interactions have an additional anisotropic 
component, as may result, {\em e.g.}, from local packing effects.
It is included perturbatively by adding an orientation dependent term 
\begin{equation}
\beta {\cal V}_{ani} = l_0 A_0 \int \! d\vec{r} \: d\vec{u} \: d\vec{u}{}'\;
\rho(\vec{r},\vec{u}) \; \rho(\vec{r},\vec{u}{}') \;
v \; \frac{5}{16 \pi} \; (3 (\vec{u} \vec{u}{}')^2 -1).
\end{equation}
The parameter $v$ is again adjustable and describes the anisotropy
per segment of the chains. 
Within this model, one dimensional self-consistent field profiles 
were calculated in the direction $z$ perpendicular to the surface, 
hence the possibility of lateral order (positional order) was not
taken into account. Fig. \ref{prof-13.7} shows some typical density
profiles. Numerous other quantities were also evaluated, such
as profiles of the nematic tensor, and the in-plane alignment
of segments $d_{\parallel}^2= \langle u_x \rangle^2 + \langle u_y \rangle^2$,
which is only nonzero when the symmetry of the $xy$ plane is broken.
The free energy was evaluated according to eqn. (\ref{fcan}), which
allows to calculate phase diagrams by means of a  Maxwell construction.

\begin{figure}

\fig{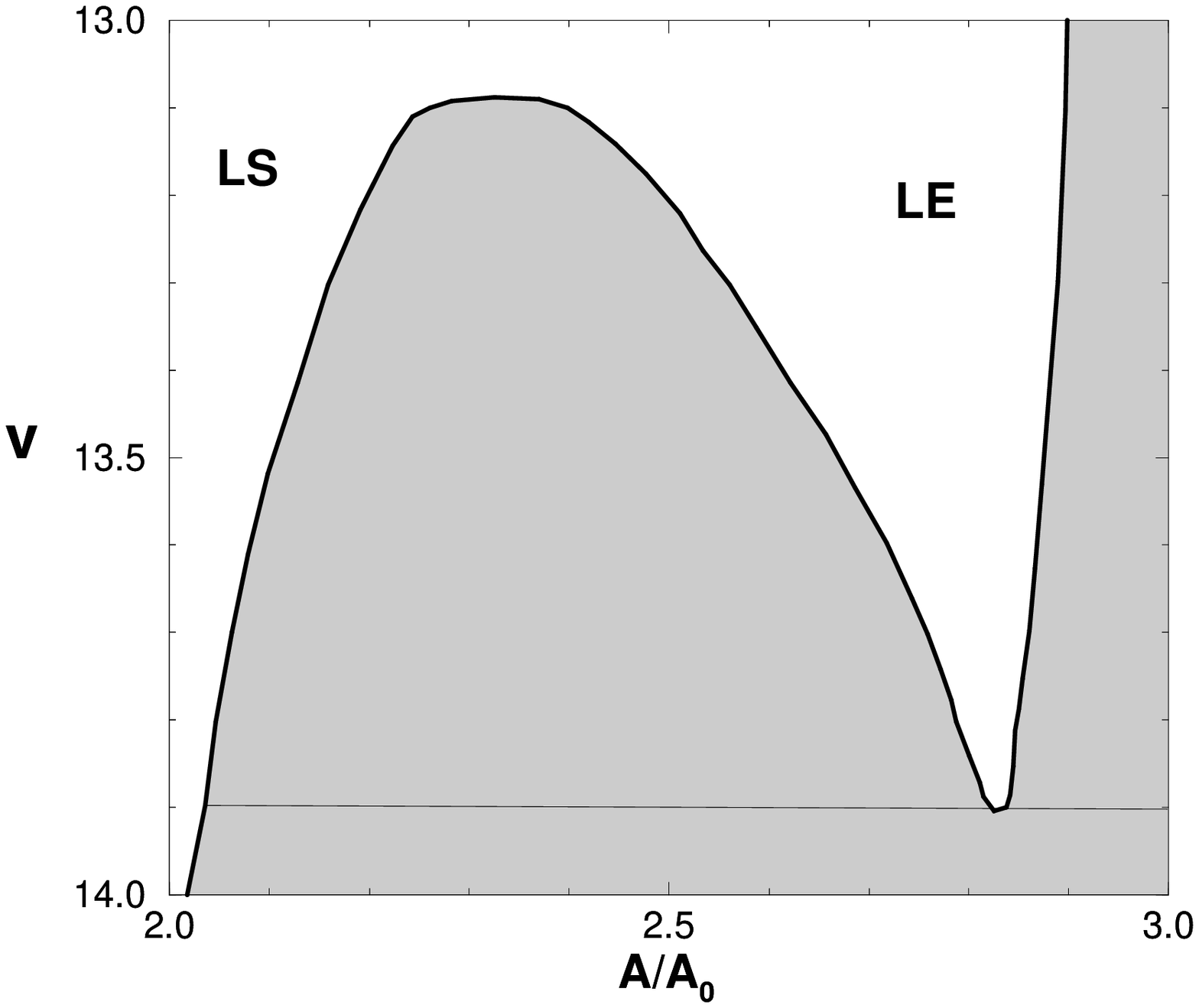}{100}{80}

\caption{\label{cut-u=2}}
Phase diagram for Langmuir monolayers for stiffer chains ($u=2$) in the 
plane of anisotropy per segment $v$ vs. area per molecule $A$ in units of 
chain diameter $A_0$. Dashed area indicates region of two phase coexistence. 
The gas phase is found at much larger values of $A$. 
Almost the same phase diagram is obtained when varying the stiffness 
parameter $u$ at fixed anisotropy $v = 13.5$.
(Critical point at $u_c=1.95$, triple point at $u_t = 2.05$ in
arbitrary units)

From Ref. \cite{FS2}.
\end{figure}

One finds that such a model indeed displays coexistence between two untilted
fluid phases. The phase behavior is driven by the stiffness $u$ 
of the chains\cite{FS4}, or alternatively by the anisotropic
interactions $v$\cite{FS2}.
On decreasing the interactions or increasing the chain flexibility, the
condensed phase merges with the expanded phase in a critical point. On 
making the chains stiffer or the interactions higher, the expanded phase
becomes unstable. Hence the coexistence region ends in a triple point and 
a critical point, like in experimental Langmuir monolayers 
(Fig. \ref{cut-u=2}). In addition, one finds
a phase with uniform tilt in one direction,
which is however metastable and buried in the coexistence region (not shown).
Since both the effect of segment interactions and 
the chain stiffness go down with increasing temperature, the $u$ axis or 
$v$ axis can be interpreted as temperature axis. 

The nature of the phase transition can thus be analyzed. The expanded
phase is stabilized by the chain entropy of sufficiently flexible
chains. The anisotropic interactions between segments, which have a 
stronger effect in systems stiffer chains, are necessary to bring about a 
distinct condensed phase. The phase transition is driven by the
interplay of the entropy of the chains and their tendency of
parallel alignment. 

\begin{figure}

\fig{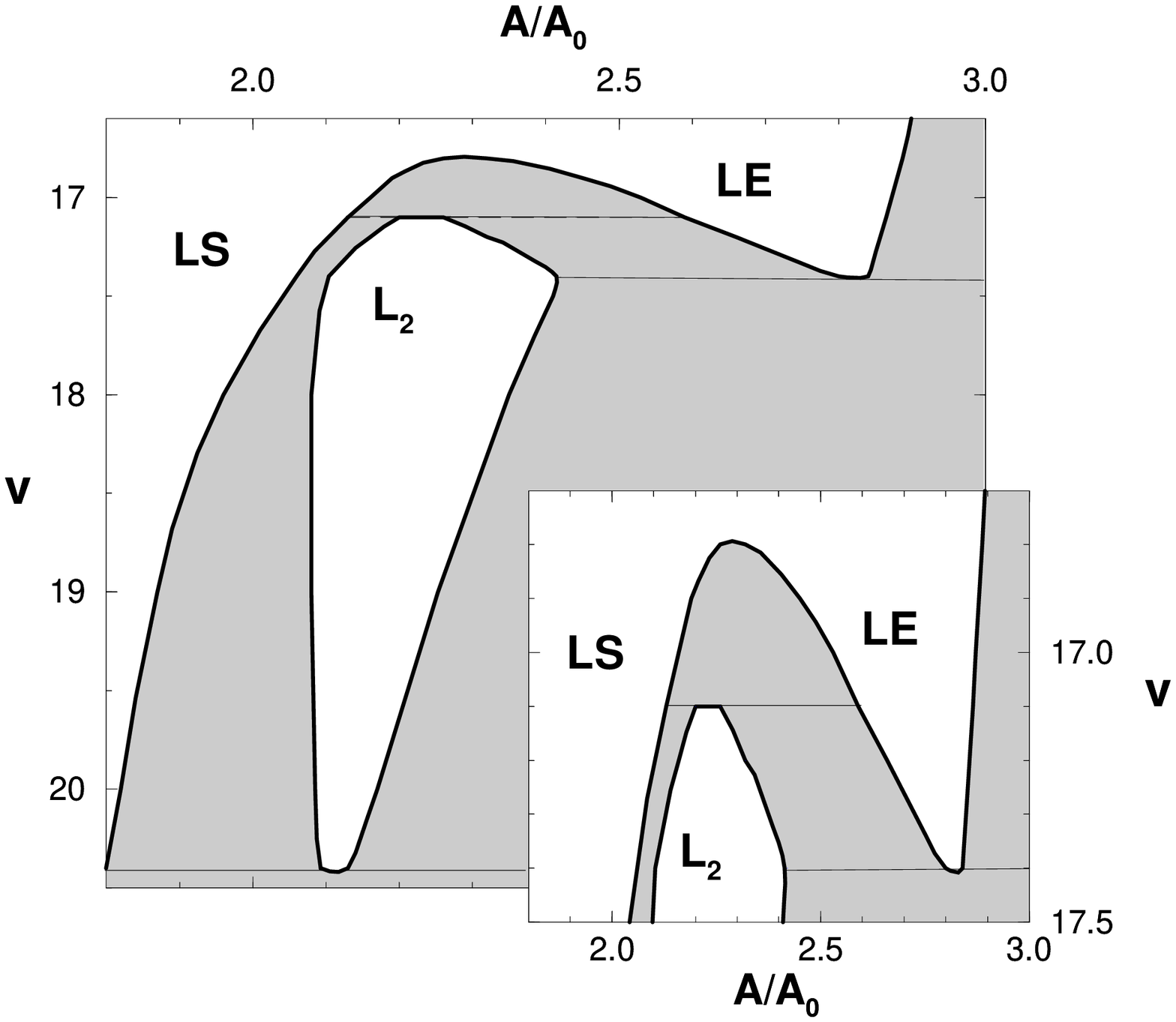}{100}{80}

\caption{\label{cut-u=1.5}}
Phase diagram for Langmuir monolayers for flexible chains ($u=1.5$) in the 
plane of anisotropy per segment $v$ vs. area per molecule $A$ in units of 
chain diameter $A_0$. The $L{}_2$ region corresponds to a phase
with broken symmetry in the $xy$ plane, where the chains are collectively
tilted in one direction.

From Ref. \cite{FS2}.
\end{figure}

Next one may ask what happens to the phase diagram (Fig. \ref{cut-u=2}) if 
the stiffness $u$ is changed substantially. From the previous results, 
we can infer that the transition values for $v$ at the critical and triple 
point are shifted in the opposite direction.
In the limit of very stiff chains, the two transitions merge, and one is 
left with only one first order transition, from the gas phase directly into 
the untilted condensed phase. This agrees with the experimental results 
of Ref. \cite{RICE}. 
If the chains are made more flexible, on the other hand, the
already mentioned tilted phase emerges as additional stable phase:
The gain of conformational entropy at the corresponding surface
densities compensates in part for the loss of surface energy per chain,
and surface coverages are stabilized which support collective tilt.
This is illustrated in Fig. \ref{cut-u=1.5}. The liquid expanded phase
can then coexist with either an untilted condensed phase or a tilted 
condensed phase. At even higher chain flexibility, the coexisting
condensed phase is entirely tilted, and turns into an untilted phase
{\em via} a continuous transition upon further compression of the monolayer 
(not shown). Hence tilted phases are stabilized by
chain flexibility.  Note however that other tilting mechanisms
are possible ({\em e.g.}, resulting from a mismatch between head size and 
chain diameter), which are presumably predominant in real 
monolayers.

In sum, the self-consistent field analysis of this model lays open
a rich and complex phenomenology. The different phases at low surface 
coverage are largely recovered. This demonstrates again the power of the 
method even for systems with relatively short chains, even though, unlike in
polymers, the predictions cannot be expected to be quantitative here.

\section{Outlook}

We have reviewed some recent advances in the self-consistent field
approach, with the aim to give a flavor of the potential and the
limitations of this method in the study of complex fluids.
Many paths of further developments are possible. 
For example, the combination of self-consistent field theories with other 
more microscopic mean field approaches, like the P-RISM theory\cite{NATH1}, 
might open promising routes to tackle new topical problems like polymer 
crystallization. 
In general, the investigation of interrelations between different
mean field approaches contributes to a deepened understanding of the 
individual methods\cite{FREED4,DONLEY}. 
Another challenging problem for the future is the
formulation of a general self-consistent field theory for semidilute 
self-avoiding chains, which would bridge between the self-avoiding
chain statistics on small length scales and the random walk chain 
statistics on larger length scales 
(Ref. \cite{WEINHOLD} is a first attempt in this direction). 
Such a method would allow, for example, to study interfaces between 
hydrophobic and hydrophilic polymer components in aqueous environment,
which are of high interest in biology.

\section*{Acknowledgements}

I wish to thank M. Schick, A. Werner, M. M\"uller, P. Janert, and K. Binder 
for fruitful and enjoyable collaborations, and M.W. Matsen, P. Nielaba 
for stimulating discussions. Partial financial support by the
Deutsche Forschungsgemeinschaft (grants Schm 985/1-1 and Bi 314/3-4),
by the Materialwissenschaftliches Forschungszentrum Mainz (MWFZ), 
and by the Graduiertenkolleg on supramolecular systems in Mainz
is acknowledged.

\clearpage

\clearpage

\clearpage

\clearpage

\clearpage

\clearpage

\clearpage

\clearpage

\clearpage

%
%
%
%

\clearpage

\clearpage

\begin{thebibliography}{99}
\bibitem{FN1}
  In some cases, it may be more convenient to work with the
  indirect correlation function 
  $y(\vec{r})=\exp[\beta W_0(\vec{r}) \gamma(\vec{r})$,
  which has the advantage of being continuous even in hard core systems,
  and to replace $U_{\alpha \beta}(\vec{r})$ in eqn. (\ref{uu}) by
  $\tilde{U}_{\alpha \beta}(\vec{r}) = 
  \exp[-\beta \: W_0(\vec{r})] -
  exp[-\beta \: W_{\alpha \beta}(\vec{r})]
  $.
\bibitem{FN2}
Here the broadening parameter $s^2$ was used as adjustable parameter
($s^2=2.5$). The actual size of the simulation box was 
$L\times L \times D$ with $L=512$ and $D=64$. 
However, the capillary wave spectrum here is neither limited
by $L$ nor by $D$, but by the time needed for capillary waves to build up. 
In systems with purely diffusive dynamics, the relaxation time for long 
wavelength capillary waves slows down dramatically with $1/q$,
($t \propto q^{-4}$ \cite{MARCUS7}).
This effectively cuts them off at some value $q_{\mbox{\tiny min}}$,
which depends only weakly on the total simulation period after an initial
relaxation time, and which is found to be 
$q_{\mbox{\tiny min}}^{-1}  \approx 60$ in our case.
For the detailed study of capillary wave effects in Ref. \cite{ANDREAS2},
smaller systems had to be studied, and nonlocal Monte Carlo moves were used.
\end{thebibliography}
\end{document}